\documentclass{vldb}

\usepackage{times,amssymb,amsmath,epsfig}
\usepackage{graphicx}
\usepackage{algorithmic}
\usepackage{algorithm}
\usepackage{comment}
\usepackage{url}

\newtheorem{definition}{Definition}
\newtheorem{theorem}{Theorem}
\newtheorem{lemma}{Lemma}
\newtheorem{example}{Example}

\begin{document}

\title{On the Relative Trust between Inconsistent Data and Inaccurate Constraints}

\author{
{George Beskales{\small $~^{1}$}~~~~~~~Ihab F. Ilyas{\small $~^{1}$}~~~~~~~Lukasz Golab{\small $~^{2}$}~~~~~~~Artur Galiullin{\small $~^{2}$}}
\vspace{1.6mm}\\
\fontsize{10}{10}\selectfont\itshape
\hspace{-0.5in} $^1$Qatar Computing Research Institute \hspace{0.8in} $^2$University of Waterloo
\vspace{0.2mm}\\
\fontsize{9}{9}\selectfont\ttfamily\upshape
\hspace{0.3in}\{gbeskales,ikaldas@\}qf.org.qa \hspace{0.3in} \{lgolab,agaliullin\}@uwaterloo.ca
}

\maketitle

\begin{abstract}

Functional dependencies (FDs) specify the intended data semantics while violations of FDs indicate deviation from these semantics.  In this paper, we study a data cleaning problem in which the FDs may not be completely correct, e.g., due to data evolution or incomplete knowledge of the data semantics.  We argue that the notion of relative trust is a crucial aspect of this problem: if the FDs are outdated, we should modify them to fit the data, but if we suspect that there are problems with the data, we should modify the data to fit the FDs.  In practice, it is usually unclear how much to trust the data versus the FDs.  To address this problem, we propose an algorithm for generating non-redundant solutions (i.e., simultaneous modifications of the data and the FDs) corresponding to various levels of relative trust.  This can help users determine the best way to modify their data and/or FDs to achieve consistency.

\end{abstract}

\section{Introduction}
\label{sec:intro}

Poor data quality is a serious and costly problem, often addressed by specifying the intended semantics using constraints such as Functional Dependencies (FDs), and modifying or discarding inconsistent data to satisfy the provided constraints.  For example, many techniques exist for editing the data in a non-redundant way so that a supplied set of FDs is satisfied \cite{BeskalesIG10,BohannonFFR05,KolahiL09}.  However, in practice, it is often unclear whether the data are incorrect or whether the intended semantics are incorrect (or both).  It is difficult to get the semantics right, especially in complex data-intensive applications, and the semantics (and/or the schema) may change over time.  Thus, practical data cleaning approaches must consider errors in the data as well as errors in the specified constraints, as illustrated by the following example.

\begin{figure}
  \center
  \includegraphics[width=3.4in]{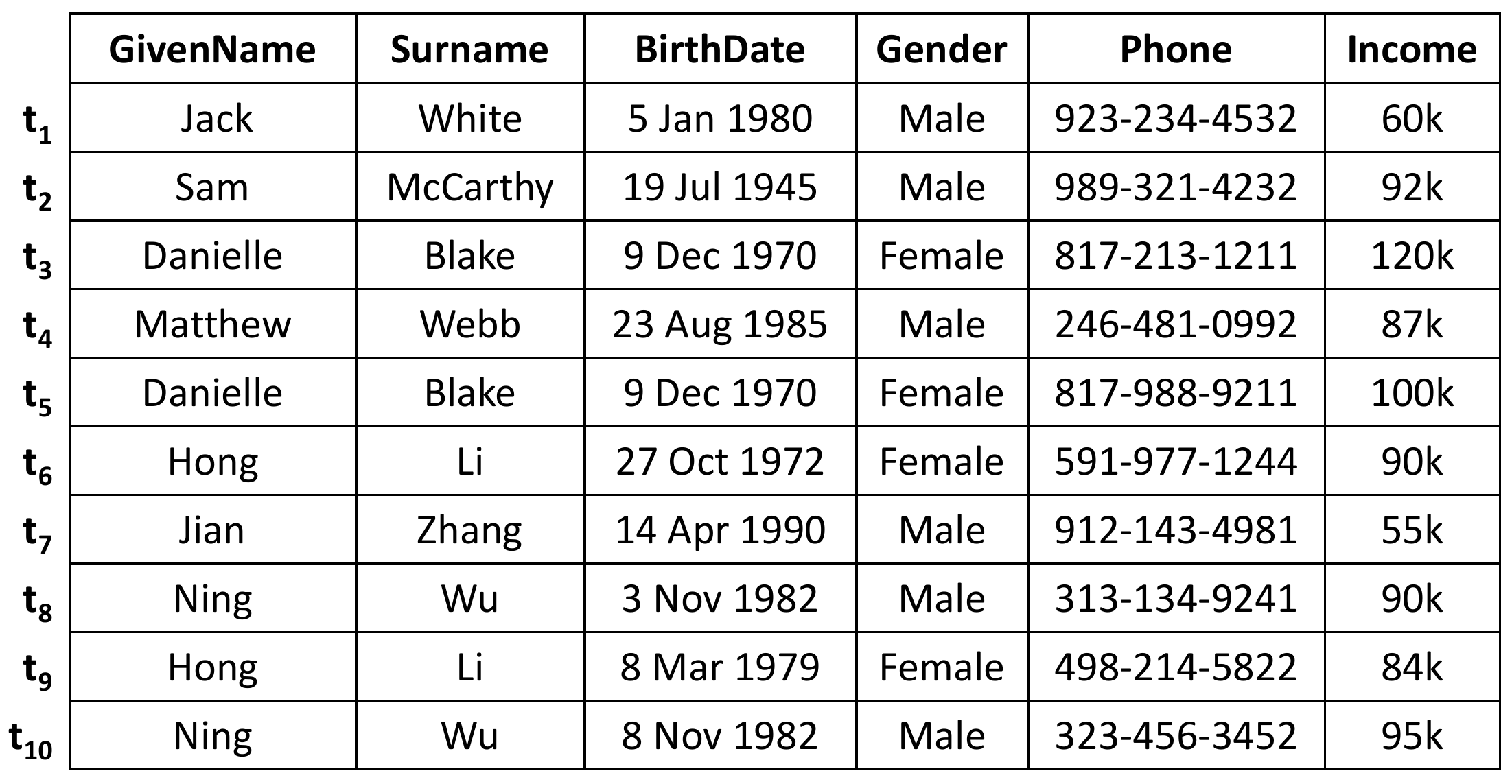}\\
  \caption{\small{An Example of a database instance of persons}}\label{fig:persons}
\end{figure}

\begin{example}
\label{ex:persons}
Figure~\ref{fig:persons} depicts a relation that holds employee information. Data are collected from various sources (e.g., Payroll records, HR) and thus might contain inconsistencies due to, for instance, duplicate records and human errors. Suppose that we initially assert the FD  \texttt{Surname, GivenName $\rightarrow$ Income}. That is, whenever two tuples agree on attributes \texttt{Surname} and \texttt{GivenName}, they must agree on \texttt{Income}.  This FD may hold for Western names, in which surname and given name may uniquely identify a person, but not for Chinese names (e.g., tuples $t_6$ and $t_9$ probably refer to different people).  Thus, we could change the FD to \texttt{Surname, GivenName, BirthDate $\rightarrow$ Income} and resolve the remaining inconsistencies by modifying the data, i.e., setting the Income of $t_5$ (or $t_3$) to be equal to that of $t_3$ (resp. $t_5$).
\end{example}

In  \cite{ChiangM11}, an algorithm was proposed to generate a single repair of both the data and the FDs, which has the lowest cost according to a unified cost model of modifying a tuple and modifying an FD (i.e., adding an extra column to its left-hand-side).  However, in practice, the data and the FDs are not always equally ``trustworthy''.  For example, FDs that were automatically discovered from legacy data may be less reliable than those manually specified by a domain expert.  Also, the reliability of data depends on many factors such as the data sources and extraction methods.  Returning to Example~\ref{ex:persons}, trusting the FD more than data suggests changing the \texttt{Income} of tuples $t_5$, $t_6$ and $t_{10}$ to be equal to the income of $t_3$, $t_9$ and $t_8$, respectively, while keeping the FD unchanged.  Trusting the data more the than FD implies modifying the FD to be \texttt{Surname, GivenName, Birthdate, Phone $\rightarrow$ Income}, while keeping the data unchanged.

In this paper, we argue that the notion of \emph{relative trust} between the data and the FDs must be taken into account when deciding how to resolve inconsistencies.  We propose an algorithm that generates multiple suggestions for how to modify the data and/or the FDs in a minimal and non-redundant way, corresponding to various levels of relative trust.  These suggested repairs can help users and domain experts determine the best way to resolve inconsistencies.

Returning to Example~\ref{ex:persons}, it is not clear whether we should modify the data alone, or add \texttt{Birthdate} (or also \texttt{Phone}) to the left-hand-side of the FD and resolve any remaining violations by modifying the data.  Without complete knowledge of the data semantics, these possibilities may not be obvious by manually inspecting the data and the FDs.  Computing various alternative fixes corresponding to different relative trust levels can give users a better idea of what could be wrong with their data and their constraints, and how to resolve problems.

Implementing the proposed relative-trust-aware approach requires solving the following technical challenges:

\begin{itemize}

\item \textbf{Minimality of changes}.
As in previous work on data repair  \cite{BeskalesIG10,BohannonFFR05,CongFGJM07,KolahiL09}, the suggested modifications to the data and the FDs should be (approximately) minimal and should avoid making redundant modifications to be meaningful.  However, it is not obvious how to define non-redundancy and minimality when both the data and the FDs can be modified, especially if we want to produce multiple suggestions, not just a single one that is globally minimal according to a unified cost model.  Furthermore, finding a data repair with the fewest possible changes is already NP-hard even if the FDs cannot be modified.


\item \textbf{Specifying Relative Trust}.
In previous work on simultaneously repairing the data and FDs  \cite{ChiangM11}, the level of relative trust was fixed and implicitly encoded in the unified cost model.  Since we want to produce multiple suggested repairs corresponding to various levels of relative trust, we need to define a semantically meaningful metric for measuring relative trust.


\end{itemize}

In this paper, we address a data cleaning problem in which we are given a set of FDs and a data set that does not comply with the specified FDs, and we return multiple non-redundant suggestions (corresponding to different levels of relative trust) for how to modify the data and/or the FDs in order to achieve consistency.  We make the following technical contributions:

\begin{itemize}

\item
We propose a simple definition of relative trust, in which a parameter $\tau$ specifies the maximum number of allowed data changes;  the smaller the $\tau$ the greater the trust in the data.  Using the notion of relative trust, we define a space of minimal FD and data repairs based on dominance with respect to $\tau$ and the amount of changes to the FDs. 

\item
We give an efficient and effective algorithm for finding a minimal modification of a set of FDs such that no more than $\tau$ data modifications will be required to satisfy the modified FDs.  The algorithm prunes the space of possible FD modifications using A* search combined with a novel heuristic that estimates the distance to an optimal set of FD modifications.  Intuitively, this algorithm computes a minimal repair of the FDs for the relative trust level specified by $\tau$.  We then give an algorithm that lists the required data modifications.  Since computing the fewest data changes required to satisfy a set of FDs is NP-hard, we resort to approximation.  The suggested repairs generated by our algorithm are \emph{provably} close to our minimality criteria on the data changes.  The approximation factor only depends on the number of FDs and the number of attributes.

\item Using the above algorithm as a subroutine, we give an algorithm for generating a sample of suggested repairs corresponding to a range of relative trust values.  We optimize this technique by reusing repairs for higher values of $\tau$ to obtain repairs for smaller $\tau$.

\end{itemize}




Finally, we perform various experiments that justify the need to incorporate relative trust in the data cleaning process and we show order-of-magnitude performance improvements of the proposed algorithms over straightforward approaches.

The remainder of the paper is organized as follows. Section~\ref{sec:notations} gives the notation and definitions used in the paper.  In Section~\ref{sec:dataFDs}, we introduce the concepts of minimal repairs and relative trust.  Section~\ref{sec:holistic}, introduces our approach to finding a nearly-minimal repair for a given relative trust value, followed by a detailed discussion of modifying the FDs in Section~\ref{sec:cheapestFDs} and modifying the data in Section~\ref{sec:datarep}.  Section~\ref{sec:range} presents the algorithm for efficiently generating multiple suggested repairs.  Section~\ref{sec:expr} presents our experimental results, Section~\ref{sec:related} discusses related work, and Section~\ref{sec:conclusion} concludes the paper.

\section{Preliminaries}
\label{sec:notations}

Let $R$ be a relation schema consisting of $m$ attributes, denoted $\{A_1,\dots,A_m\}$. We denote by $|R|$ the number of attributes in $R$.  $Dom(A)$ denotes the domain of an attribute $A \in R$. We assume that attribute domains are unbounded. An instance $I$ of $R$ is a set of tuples, each of which belongs to the domain $Dom(A_1) \times \dots \times Dom(A_m)$. 
We refer to an attribute $A \in R$ of a tuple $t \in I$ as a \emph{cell}, denoted $t[A]$. 

For two attribute sets $X,Y \subseteq R$, a functional dependency (FD) $X \rightarrow Y$ holds on an instance $I$, denoted
$I \models  X \rightarrow Y$, iff for every two tuples $t_1,t_2$ in $I$, $t_1[X] = t_2[X]$ implies $t_1[Y] =t_2[Y]$. Let $\Sigma$ be the set of FDs defined over $R$. We denote by $|\Sigma|$ the number of FDs in $\Sigma$. We say that $I$ satisfies $\Sigma$, written $I \models \Sigma$, iff the tuples in $I$ do not violate any FD in $\Sigma$. We assume that $\Sigma$ is minimal \cite{AbiteboulHV95}, and each FD is of the form $X \rightarrow A$, where $X \subset R$ and $A \in R$.

We use the notion of V-instances, which was first introduced in \cite{KolahiL09}, to concisely represent multiple data instances. In V-instances, cells can be set to variables that may be instantiated in a specific way.

\begin{definition} \label{def:vinstance}
\textbf{V-instance}.
Given a set of variables $\{v_1^A,v_2^A,\dots\}$ for each attribute $A \in R$, a V-instance $I$ of $R$ is an instance of $R$ where each cell $t[A]$ in $I$ can be assigned to either a constant in $Dom(A)$, or a variable $v^A_i$.
\end{definition}

A V-instance $I$ represents multiple (ground) instances of $R$ that can be obtained by assigning each variable $v^A_i$ to any value from $Dom(A)$ that is not among the $A$-values already occurring in $I$, and such that no two distinct variables $v_i^A$ and $v_j^A$ can have equal values. For brevity, we refer to a V-instance as an instance in the remainder of the paper.

We say that a vector $X=(x_1,\dots,x_k)$ dominates another vector $Y=(y_1,\dots,y_k)$, written $X \prec Y$, iff for $i\in \{1,\dots,k\}$, $x_i \leq y_i$, and at least one element $x_j$ in $X$ is strictly less than the corresponding element $y_j$ in $Y$.

\section{Spaces of Possible Repairs}
\label{sec:dataFDs}

In this section, we define a space of minimal repairs of both data and FDs (Section~\ref{sec:minrep}) and we present our notion of relative trust (Section~\ref{sec:param}).

\subsection{Minimal Repairs of Data and FDs}
\label{sec:minrep}


We consider data repairs that change cells in $I$ rather than deleting tuples from $I$. We denote by $\mathcal{S}(I)$ all possible repairs of $I$. All instances in $\mathcal{S}(I)$ have the same number of tuples as $I$. Because we aim at modifying a given set of FDs, rather than discovering a new set of FDs from scratch, we restrict the allowed FD modifications to those that relax (i.e., weaken) the supplied FDs.  We do not consider adding new constraints.  That is, $\Sigma'$ is a possible modification of $\Sigma$ iff $I \models \Sigma$ implies $I\models \Sigma'$, for any data instance $I$. Given a set of FDs $\Sigma$, we denote by $\mathcal{S}(\Sigma)$ the set of all possible modifications of $\Sigma$ resulting from relaxing the FDs in $\Sigma$ in all possible ways. We define the universe of possible repairs as follows.

\begin{definition} \textbf{Universe of Data and FDs Repairs.}
Given a data instance $I$ and a set of FDs $\Sigma$, the universe of repairs of data and FDs, denoted ${\bf U}$, is the set of all possible pairs $(\Sigma',I')$ such that $\Sigma' \in \mathcal{S}(\Sigma)$, $I' \in\mathcal{S}(I)$, and $I' \models \Sigma'$.
\end{definition}

We focus on a subset of $\textbf{U}$ that are Pareto-optimal with respect to two distance functions: $dist_c(\Sigma,\Sigma')$ that measures the distance between two sets of FDs, and $dist_d(I,I')$ that measures the distance between two database instances. We refer to such repairs as \emph{minimal repairs}, defined as follows.




\begin{definition} \textbf{Minimal Repair.}
Given an instance $I$ and a set of FDs $\Sigma$, a repair $(\Sigma',I') \in {\bf U}$ is minimal iff $\nexists (\Sigma'',I'')\in {\bf U}$ such that $(dist_c(\Sigma,\Sigma''),dist_d(I,I'')) \prec (dist_c(\Sigma,\Sigma'),dist_d(I,I'))$. 
\end{definition}


We deliberately avoid aggregating changes to data and changes to FDs into one metric in order to enable using various metrics for measuring both types of changes, which might be incomparable. For example, one metric for measuring changes in $\Sigma$ is the number of modified FDs in $\Sigma$, while changes in $I$ could be measured by the number of changed cells. Also, this approach provides a wide spectrum of Pareto-optimal repairs that ranges from completely trusting $I$ (and only changing $\Sigma$) to completely trusting $\Sigma$ (and only changing $I$).

For a repair $I'$ of $I$, we denote by $\Delta_d(I,I')$ the cells that have different values in $I$ and $I'$.
We use the cardinality of $\Delta_d(I,I')$ to measure the distance between two instances, which has been widely used in previous data cleaning techniques (e.g., \cite{BohannonFFR05,CongFGJM07,KolahiL09}). That is, $dist_d(I,I') = |\Delta_d(I,I')|$.

Recall that we restrict the modifications to $\Sigma$ to those that relax the constraints in $\Sigma$.  Thus, an FD $F'$ is a possible modification of an FD $F$ iff $I \models F \Rightarrow I \models F'$, for any instance $I$. We use a simple relaxation mechanism: we only allow appending zero or more attributes to the left-hand-side (LHS) of an FD. Formally, an FD $X\rightarrow A \in \Sigma$ can be modified by appending a set of attributes $Y \subseteq (R \setminus XA)$ to the LHS, resulting in an FD $XY \rightarrow A$. We disallow adding $A$ to the LHS to prevent producing trivial FDs.

Note that different FDs in $\Sigma$ might be modified to the same FD. For example, both $A \rightarrow B$ and $C \rightarrow B$ can be modified to $AC \rightarrow B$. Therefore, the number of FDs in any $\Sigma' \in \mathcal{S}(\Sigma)$ is less than or equal to the number of FDs in $\Sigma$.  We maintain a mapping between each FD in $\Sigma$, and its corresponding repair in $\Sigma'$. 
Without loss of generality, we assume hereafter that $|\Sigma'| = |\Sigma|$ by allowing duplicate FDs in $\Sigma'$. 

We define the distance between two sets of FDs as follows.
For $\Sigma = \{X_1\rightarrow A_1,\dots, X_z \rightarrow A_z\}$ and $\Sigma' = \{Y_1 X_1 \rightarrow A_1,\dots, Y_z X_z \rightarrow A_z\}$, the term $\Delta_c(\Sigma,\Sigma')$ denotes a vector $(Y_1,\dots,Y_z)$, which consists of LHS extensions to FDs in $\Sigma$ in $\Sigma'$. To measure the distance between $\Sigma$ and $\Sigma'$, we use the function $\sum_{Y \in \Delta_c(\Sigma,\Sigma'')} w(Y)$, where $w(Y)$ is a weighting function that determines the relative penalty of adding a set of attributes $Y$. The weighting function $w(.)$ is intuitively non-negative and monotone (i.e., for any two attribute sets $X$ and $Y$, $X \subseteq Y$ implies that $w(X) \leq w(Y)$). A simple example of $w(Y)$ is the number of attributes in $Y$. However, this does not distinguish between attributes that have different characteristics. Other features of appended attributes can be used for obtaining other definitions of $w(.)$.  For example, consider two attributes $A$ and $B$ that could be appended to the LHS of an FD, where $A$ is a key (i.e., $A \rightarrow R$), while $B$ is not. Intuitively, appending $A$ should be more expensive that appending $B$  because the new FD in the former case is trivially satisfied. In general, the more informative a set of attributes is, the more expensive it is when being appended to the LHS of an FD. The information captured by a set of attributes can be measured using various metrics, such as the number of distinct values of $Y$ in $I$, and the entropy of $Y$. Another definition of $w(Y)$ could rely on the increase in the description length for modeling $I$ using FDs due to appending $Y$ (refer to \cite{ChiangM11,KramerP96}).

In general, $w(Y)$ depends on a given data instance to evaluate the weight of $Y$. Therefore, changing the cells in $I$ during the repair generating algorithm might affect the weights of attributes. We make a simplifying assumption that $w(Y)$ depends only on the initial instance $I$.  This is based on an observation that the number of violations in $I$ with respect to $\Sigma$ is typically much smaller than the size of $I$, and thus repairing data does not significantly change the characteristics of attributes such as entropy and the number of distinct values.

\subsection{Relative Trust in Data vs. FDs}
\label{sec:param}

We defined a space of minimal repairs that covers a wide spectrum, ranging from repairs that only alter the data, while keeping the FDs unchanged, to repairs that only alter the FDs, while keeping the data unchanged.  
The idea behind relative trust is to limit the maximum number of cell changes that can be performed while obtaining $I'$ to a threshold $\tau$, and to obtain a set of FDs $\Sigma'$ that is the closest to $\Sigma$ and is satisfied by $I'$. The obtained repair $(\Sigma',I')$ is called a $\tau$-constrained repair, formally defined as follows.

\begin{definition} \textbf{$\tau$-constrained Repair}
\label{def:const_min}
Given an instance $I$, a set of FDs $\Sigma$, and a threshold $\tau$, a $\tau$-constrained repair $(\Sigma',I')$ is a repair in $\bf U$ such that $dist_d(I,I') \leq \tau$, and no other repair $(\Sigma'',I'')\in {\bf U}$ has $(dist_c(\Sigma,\Sigma''),dist_d(I,I'')) \prec (dist_c(\Sigma,\Sigma'),\tau)$.
\end{definition}

In other words, a $\tau$-constrained repair is a repair in $\bf U$ whose distance to $I$ is less than or equal to $\tau$, and which has the minimum distance to $\Sigma$ across all repairs in $\bf U$ with distance to $I$ also less than or equal to $\tau$. We break ties using the distance to $I$ (i.e., if two repairs have an equal distance to $\Sigma$ and have distances to $I$ less than or equal to $\tau$, we choose the one closer to $I$).

Possible values of $\tau$ range from 0 to the minimum number of cells changes that must be applied to $I$ in order to satisfy $\Sigma$, denoted $\delta_{opt}(\Sigma,I)$. We can also specify the threshold on the number of allowed cell changes as a percentage of $\delta_{opt}(\Sigma,I)$, denoted $\tau_r$ (i.e., $\tau_r = \tau / \delta_{opt}(\Sigma,I)$). 


The mapping between minimal repairs and $\tau$-constrained repairs is as follows. (1) Each $\tau$-constrained repair is a minimal repair; (2) All minimal repairs can be found by varying the relative trust $\tau$ in the range $[0,\delta_{opt}(\Sigma,I)]$, and obtaining the corresponding $\tau$-constrained repair.  Specifically, each minimal repair $(\Sigma',I')$ is equal to a $\tau$-constrained repair, where $\tau$ is in the range defined as follows.  Let $(\Sigma'',I'')$ be the minimal repair with the smallest $dist_d(I,I'')$ that is strictly greater than $dist_d(I,I')$. If such a repair does not exist, let $(\Sigma'',I'')$ be $(\phi,\phi)$. The range of $\tau$ is defined as follows.

\begin{equation}
\label{eq:range}
\tau \in  \left\{
 \begin{array}{ll}
    $[$dist_d(I,I'),dist_d(I,I'')) & \textrm{if} \  (\Sigma'',I'') \neq (\phi,\phi)
    \\ \\
    $[$dist_d(I,I'),\infty) &  \textrm{if} \ (\Sigma'',I'') = (\phi,\phi)
\end{array} \right.
\end{equation}

If $(\Sigma'',I'') = (\phi,\phi)$, the range $[dist_d(I,I'), \infty)$ corresponds to a unique minimal repair where $dist_d(I,I')$ is equal to $\delta_{opt}(\Sigma,I)$. We prove these two points in the following theorem (proof is in Appendix~\ref{sec:proof:mapping}).

\begin{theorem}
\label{thm:mapping}
Each $\tau$-constrained repair is a minimal repair.  Each minimal repair $(\Sigma',I')$ corresponds to a $\tau$-constrained repair, where $\tau$ belongs to the range defined in Equation~\ref{eq:range}.
\end{theorem}

\section{Computing a Single Repair for a Given Relative Trust Level}
\label{sec:holistic}

There is a strong interplay between modifying the data and the FDs.  Obtaining a data instance that is closest to $I$, while satisfying a set of FDs $\Sigma'$ highly depends on $\Sigma'$. Also, obtaining a set of FDs $\Sigma'$ that is closest to $\Sigma$, such that $\Sigma'$ holds in a given data instance $I'$ highly depends on the instance $I'$.  This interplay represents the main challenge for simultaneously modifying the data and the FDs.

For example, consider a simple approach that alternates between editing the data and modifying the FDs until we reach consistency.  This may not give a minimal repair (e.g., we might make a data change in one step that turns out to be redundant after we change one of the FDs in a subsequent step). Furthermore, we may have to make more than $\tau$ cell changes because it is difficult to predict the amount of necessary data changes while modifying the FDs.

Our solution to generating a minimal repair for a given level of relative trust consists of two steps. In the first step, we modify the FDs to obtain a set $\Sigma'$ that is as close as possible to $\Sigma$, while guaranteeing that there exists a data repair $I'$ satisfying $\Sigma'$ with a distance to $I$ less than or equal to $\tau$. In the second step, we materialize the data instance $I'$ by modifying $I$ with respect to $\Sigma'$ in a minimal way. We describe this approach in Algorithm~\ref{alg:repair}.


Finding $\Sigma'$ in the first step requires computing the minimum number of cell changes in $I$ to satisfy $\Sigma'$ (i.e., $\delta_{opt}(\Sigma',I)$).
Note that computing $\delta_{opt}(\Sigma',I)$ does not require materialization of an instance $I'$ that satisfies $\Sigma'$ and have the minimum number of changes. Instead, we collect enough statistics about the violations in data to compute $\delta_{opt}(\Sigma',I)$. We will discuss this step in more detail in Section~\ref{sec:cheapestFDs}.  Obtaining a modified instance $I'$ in line 3 will be discussed in Section~\ref{sec:datarep}.

\begin{algorithm}
\small
\caption{{\texttt{Repair\_Data\_FDs($\Sigma$,$I$,$\tau$)}}}
\label{alg:repair}
\begin{algorithmic}[1]

\STATE obtain $\Sigma'$ from $\mathcal{S}(\Sigma)$ such that $\delta_{opt}(\Sigma',I) \leq \tau$, and no other $\Sigma'' \in \mathcal{S}(\Sigma)$ with $\delta_{opt}(\Sigma'',I) \leq \tau$ has $dist_c(\Sigma,\Sigma'') < dist_c(\Sigma,\Sigma')$. (ties are broken using $\delta_{opt}(\Sigma',I)$)

\IF {$\Sigma' \neq \phi$}

\STATE obtain $I'$ that satisfies $\Sigma'$ while performing at most $\delta_{opt}(\Sigma',I)$ cell changes, and return $(\Sigma',I')$.

\ELSE

\STATE Return $(\phi,\phi)$

\ENDIF

\end{algorithmic}
\end{algorithm}

The following theorem establishes the link between the repairs generated by Algorithm~\ref{alg:repair} and Definition~\ref{def:const_min}. The proof is in Appendix~\ref{sec:proof:tau_repairs}.

\begin{theorem}
\label{thm:tau_repairs}
Repairs generated by Algorithm~\ref{alg:repair} are $\tau$-constrained repairs.
\end{theorem}

A key step in Algorithm~\ref{alg:repair} is computing $\delta_{opt}(\Sigma',I)$ (i.e., the minimum number of cells in $I$ that have to be changed in order to satisfy $\Sigma'$). Unfortunately, computing the exact minimum number of cell changes when $\Sigma'$ contains at least two FDs is NP-hard \cite{KolahiL09}.  We will propose an approximate solution based on upper-bounding the minimum number of necessary cell changes. Assume that there exists a $P$-approximate upper bound on $\delta_{opt}(\Sigma',I)$, denoted  $\delta_{P}(\Sigma',I)$ (details are in Section~\ref{sec:datarep}). That is,  $\delta_{opt}(\Sigma',I)\leq \delta_{P}(\Sigma',I) \leq P \cdot \delta_{opt}(\Sigma',I)$, for some constant $P$. By using $\delta_{P}(\Sigma',I)$ in place of $\delta_{opt}(\Sigma',I)$ in Algorithm~\ref{alg:repair}, we can satisfy the criteria in Definition~\ref{def:const_min} in a $P$-approximate way. Specifically, the repair generated by Algorithm~\ref{alg:repair} becomes a $P$-approximate $\tau$-constrained repair, which is defined as follows (the proof is similar to Theorem~\ref{thm:tau_repairs}).



\begin{definition} \textbf{$P$-approximate $\tau$-constrained Repair}
\label{def:approx_const_min}
Given an instance $I$, a set of FDs $\Sigma$, and a threshold $\tau$, a $P$-approximate $\tau$-constrained repair $(\Sigma',I')$ is a repair in $\bf U$ such that $dist_d(I,I') \leq \tau$, and no other repair $(\Sigma'',I'')\in {\bf U}$ has $(dist_c(\Sigma,\Sigma''), P \cdot dist_d(I,I'')) \prec (dist_c(\Sigma,\Sigma'), \tau)$.
\end{definition}

%
%

In the remainder of this paper, we present an implementation of line 1 (Section~\ref{sec:cheapestFDs}) and line 3 (Section~\ref{sec:datarep}) of Algorithm~\ref{alg:repair}.  Our implementation is $P$-approximate, as defined above, with $P = 2 \cdot \min\{|R|-1,|\Sigma|\}$, where $|R|$ denotes the number of attributes in $R$, and $|\Sigma|$ denotes the number of FDs in $\Sigma$.


\section{Minimally Modifying the FDs}
\label{sec:cheapestFDs}


In this section, we show how to obtain a modified set of FDs $\Sigma'$ that is part of a $P$-approximate $\tau$-constrained repair (line 1 of Algorithm~\ref{alg:repair}). That is, we need to obtain $\Sigma' \in \mathcal{S}(\Sigma)$ such that $\delta_{P}(\Sigma',I) \leq \tau$, and no other FD set $\Sigma'' \in \mathcal{S}(\Sigma)$ with $\delta_{P}(\Sigma'',I) \leq \tau$ has $dist_c(\Sigma,\Sigma'') < dist_c(\Sigma,\Sigma')$.

First, we need to introduce the notion of a conflict graph of $I$ with respect to $\Sigma$, which was previously used in \cite{ArenasBC01}:

\begin{definition}
\textbf{Conflict Graph.}
\label{def:conflict_graph}
A conflict graph of an instance $I$ and a set of  FDs $\Sigma$ is an undirected graph whose set of vertices is the set of tuples in $I$, and whose set of edges consists of all edges $(t_i,t_j)$ such that $t_i$ and $t_j$ violate at least one FD in $\Sigma$.
\end{definition}

\begin{figure}
 \center
  \includegraphics[width=2.5in]{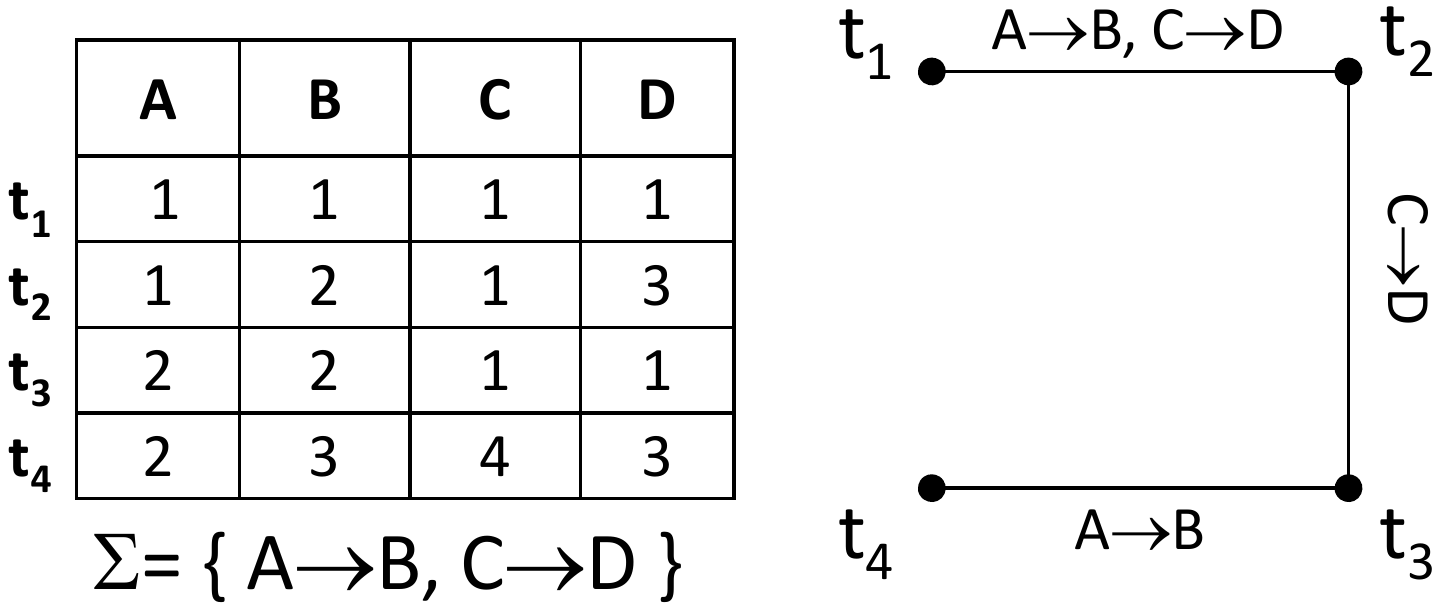}\\
    \caption{An example of a conflict graph}\label{fig:conflict_graph}
\end{figure}

Figure~\ref{fig:conflict_graph} shows an instance $I$, a set of FDs $\Sigma$, and the corresponding conflict graph. The label of each edge represents the FDs that are violated by the edge vertices.


In Section~\ref{sec:datarep}, we present an algorithm to obtain an instance repair $I'$ that satisfies a set of FDs $\Sigma' \in\mathcal{S}(\Sigma)$. The number of cell changes performed by our algorithm is linked to the conflict graph of $\Sigma'$ and $I$ as follows. Let $C_{2opt}(\Sigma',I)$ be a 2-approximate minimum vertex cover of the conflict graph of $\Sigma'$ and $I$, which we can obtain in PTIME using a greedy algorithm \cite{GJ90}. The number of cell changes performed by our algorithm is at most $\alpha \cdot  |C_{2opt}(\Sigma',I)|$, where $\alpha = \min\{|R|-1,|\Sigma|\}$. Moreover, we prove that the number of changed cells is $2\alpha$-approximately minimal. Therefore, we define $\delta_P(\Sigma',I)$ as $\alpha \cdot |C_{2opt}(\Sigma',I)|$, which represents a $2\alpha$-approximate upper bound of $\delta_{opt}(\Sigma',I)$ that can be computed in PTIME. Based on the definition of $\delta_P(\Sigma',I)$, our goal in this section can be rewritten as follows:  obtain $\Sigma' \in \mathcal{S}(\Sigma)$ such that $C_{2opt}(\Sigma',I) \leq \frac{\tau}{\alpha}$, and no other FD set $\Sigma'' \in \mathcal{S}(\Sigma)$ with $C_{2opt}(\Sigma'',I) \leq \frac{\tau}{\alpha}$ has $dist_c(\Sigma,\Sigma'') < dist_c(\Sigma,\Sigma')$.

\begin{figure}[t]
 \center
  \includegraphics[width=3.4in]{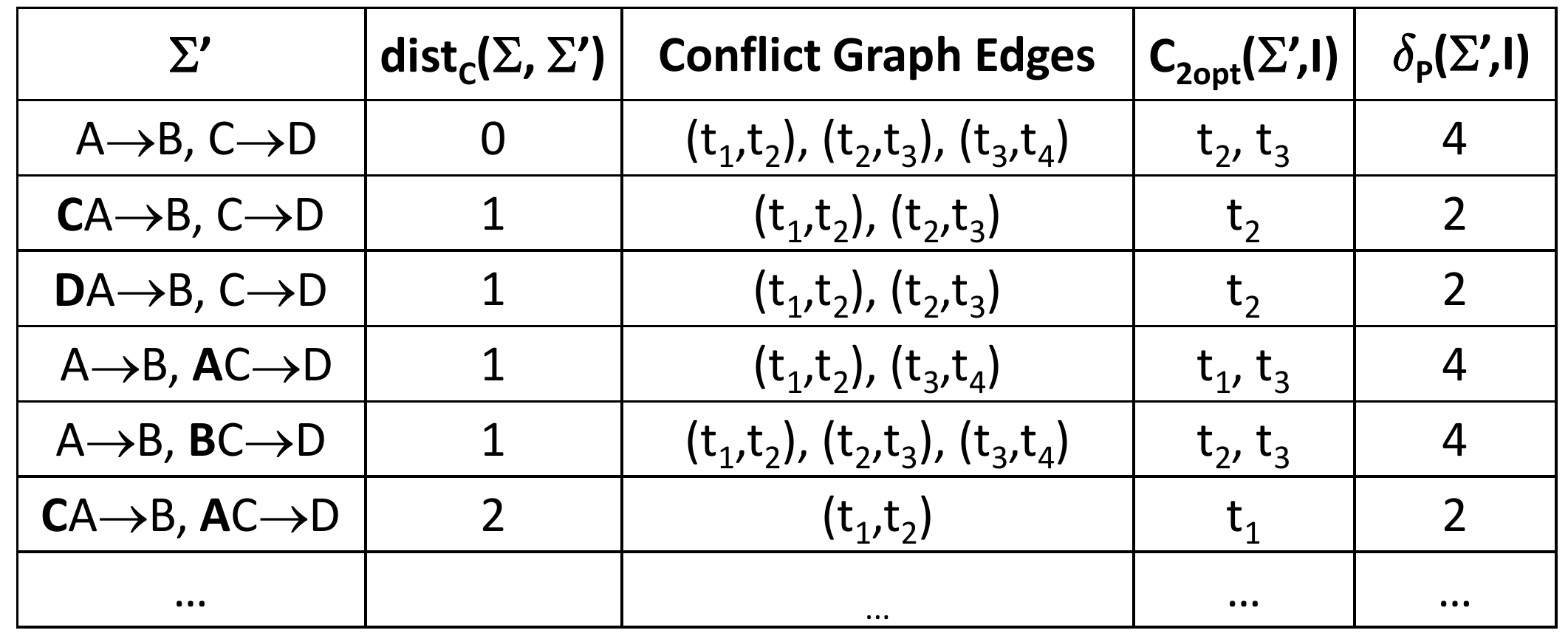}\\
  \caption{An example of multiple FD repairs}\label{fig:sigmas}
\end{figure}

Figure~\ref{fig:sigmas} depicts several possible modifications of $\Sigma$ from Figure~\ref{fig:conflict_graph}, along with $dist_c(\Sigma,\Sigma')$ (assuming that the weighting function $w(Y)$ is equal to $|Y|$), the corresponding conflict graph, $C_{2opt}(\Sigma',I)$, and $\delta_P(\Sigma',I)$. For $\tau = 2$, the modifications of $\Sigma$ that are part of $P$-approximate $\tau$-constrained repairs are $\{CA \rightarrow B, C \rightarrow D\}$ and $\{DA \rightarrow B, C \rightarrow D\}$.

\subsection{Searching the Space of FD Modifications}

We model the possible FD modifications $\mathcal{S}(\Sigma)$ as a state space, where for each $\Sigma' \in \mathcal{S}(\Sigma)$, there exists a state representing $\Delta_c(\Sigma,\Sigma')$ (i.e., the vector of attribute sets appended to LHSs of FDs to obtain $\Sigma'$). Additionally, we call $\Delta_c(\Sigma,\Sigma')$ a \emph{goal state} iff $\delta_P(\Sigma',I) \leq \tau$, for a given threshold value $\tau$ (or equivalently, $C_{2opt}(\Sigma',I) \leq \frac{\tau}{\alpha}$).  The cost of a state $\Delta_c(\Sigma,\Sigma')$ is equal to $dist_c(\Sigma,\Sigma')$. We assume that the weighting function $w(.)$ is monotone and non-negative. Our goal is to locate the cheapest goal state for a given value of $\tau$, which amounts to finding an FD set $\Sigma'$ that is part of a $P$-approximate $\tau$-constrained repair.

The monotonicity of the weighting function $w$ (and hence the monotonicity of the overall cost function) allows for pruning a large part of the state space. We say that a state $(Y_1,\dots,Y_z)$ \emph{extends} another state $(Y'_1,\dots,Y'_z)$, where $z = |\Sigma|$, iff for all $i \in \{1,\dots,z\}$,  $Y'_i \subseteq Y_i$. Clearly, if $(Y_1,\dots,Y_z)$ is a goal state, we can prune all the FD sets that extend it because $w(.)$ is monotone.

\begin{figure}[t]
  \center
  \includegraphics[width=3.4in]{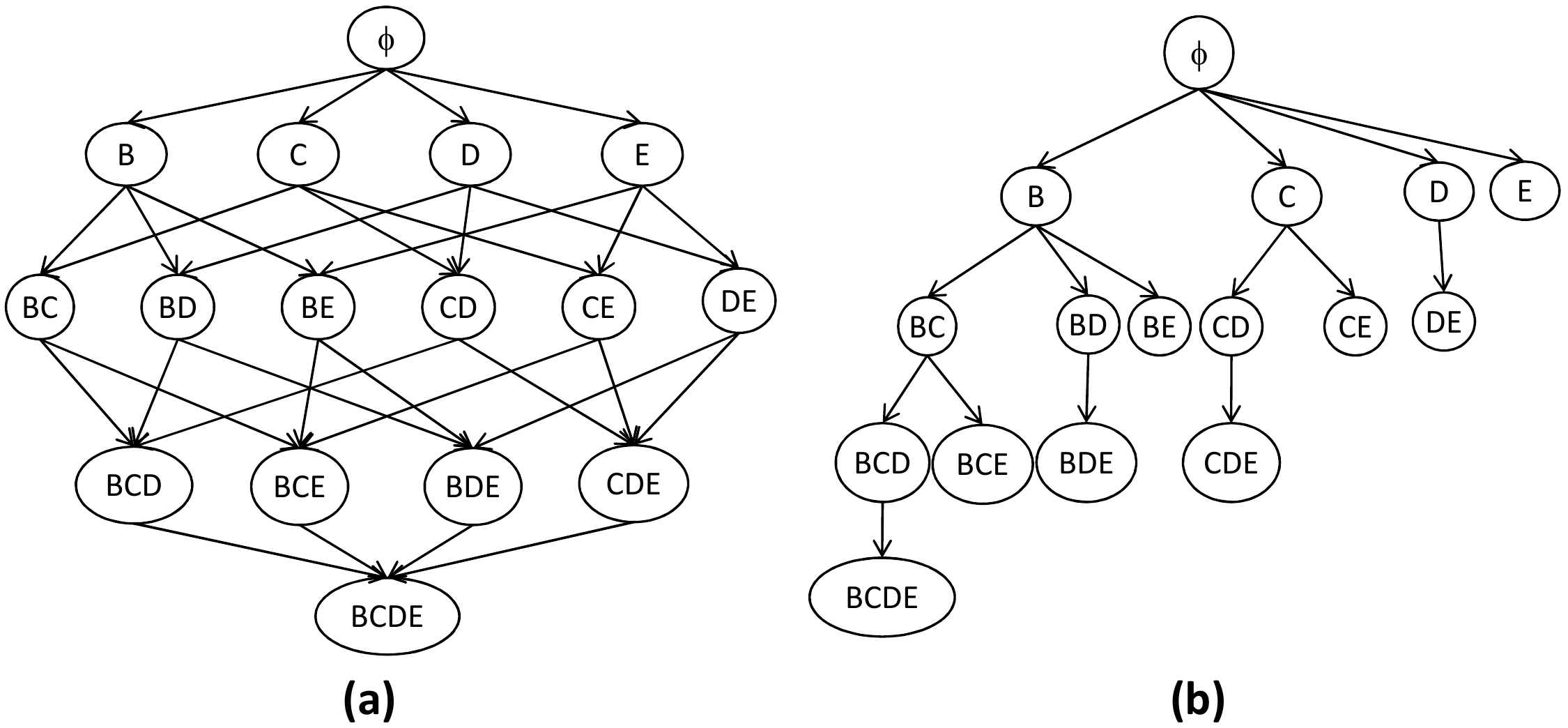}\\
  \caption{The state search space for $R=\{A,B,C,D,E,F\}$ and $\Sigma=\{A \rightarrow F\}$ (a) a graph search space (b) a tree
  search space}\label{fig:latticetree}
\end{figure}

\begin{figure}
  \center
  \includegraphics[width=2.7in]{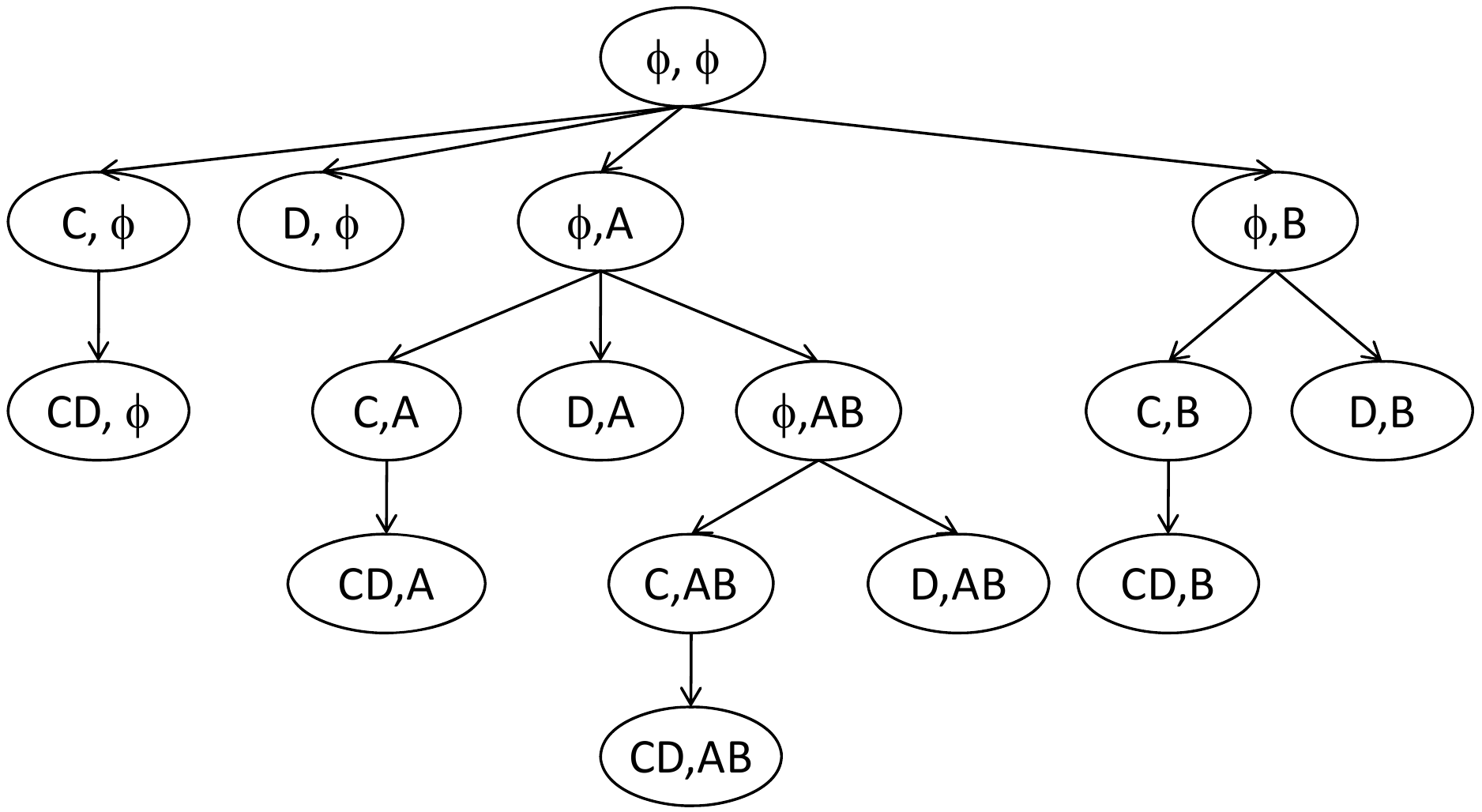}\\
  \caption{A space for $R=\{A,B,C,D\}$ and $\Sigma=\{A \rightarrow B, C \rightarrow D\}$}\label{fig:tree2}
\end{figure}

In Figure~\ref{fig:latticetree}(a), we show all the states for $R=\{A,B,C,D,E,F\}$ and $\Sigma=\{A \rightarrow F\}$. Each arrow in Figure~\ref{fig:latticetree}(a) indicates that the destination state extends the source state by adding exactly one attribute. We can find the cheapest goal state by traversing the graph in Figure~\ref{fig:latticetree}(a). For example, we can use a level-wise breadth-first search strategy \cite{PEARL89}, which iterates over states with the same number of attributes, and, for each such set of states, we determine whether any state is a goal state. If one or more goal states are found at the current level, we return the cheapest goal state and terminate the search.

We can optimize the search by adopting best-first traversal of the states graph \cite{PEARL89}.  That is, we maintain a list of states to be visited next, called the \emph{open list}, which initially contains the state $(\phi,\dots,\phi)$, and a list of states that have been visited, called the \emph{closed list}. In each iteration, we pick the cheapest state $S$ from the open list, and test whether $S$ is a goal state. If $S$ is a goal state, we return it and terminate the search. Otherwise, we add $S$ to the closed list, and we insert into the open list all the states that extend $S$ by exactly one attribute and are not in the closed list.

We can avoid using a closed list that keeps track of visited states, and hence reduce the running time, by ensuring that each state can only be reached from the initial state $(\phi,\dots,\phi)$ using a unique path. In other words, we need to reduce the graph in Figure~\ref{fig:latticetree}(a) to a tree (e.g., Figure~\ref{fig:latticetree}(b)). To achieve this, we assign each state, except $(\phi,\dots,\phi)$, to a single parent. Assume that attributes in $R$ are totally ordered (e.g., lexicographically). For $\Sigma$ with a single FD, the parent of a state $Y$ is another state $Y\setminus \{A\}$  where $A$ is the greatest attribute in $Y$. Figure~\ref{fig:latticetree}(b) shows the search tree  that is equivalent to the search graph in Figure~\ref{fig:latticetree}(a). In general, when $\Sigma$ contains multiple FDs, the parent of a state $(Y_1,\dots,Y_z)$ is determined as follows. Let $A$ be the greatest attribute in $\bigcup_{i=1}^z Y_i$, and $j$ be the index of the last element in the vector $(Y_1,\dots,Y_z)$ that contains $A$. The parent of the state $(Y_1,\dots,Y_z)$ is another state $(Y_1,\dots,Y_{j-1},Y_j \setminus \{A\},Y_{j+1},\dots,Y_z)$. Figure~\ref{fig:tree2} depicts an example search space for the two FDs shown in Figure~\ref{fig:conflict_graph}.

\subsection{A*-based Search Algorithm}
\label{sec:astar}

One problem with best-first tree traversal is that it might visit cheap states that only lead to expensive goal states or no goal states at all.  A* search \cite{PEARL89} avoids this by estimating the cost of the cheapest goal state reachable (i.e., descending) from each state $S$ in the open list, denoted $\underbar{gc}(S)$, and visiting the state with the smallest $\underbar{gc}(S)$ first. In order to maintain soundness of the algorithm (i.e., returning the cheapest goal state), we must not overestimate the cost of the cheapest goal state reachable from a state $S$ \cite{PEARL89}.

\begin{algorithm}
\small
\caption{\texttt{Modify\_FDs($\Sigma,I,\tau$)}}
\label{alg:astar}
\begin{algorithmic}[1]
\STATE construct the conflict graph $G$ of $\Sigma$ and $I$, and obtain the set of all difference sets in $G$, denoted $\mathcal{D}$

\STATE $PQ \leftarrow \{ (\phi,\dots,\phi) \}$

\WHILE {$PQ$ is not empty}

\STATE pick the state $S_h$ with the smallest value of $\underbar{gc}(.)$ from $PQ$

\STATE let $\Sigma_h$ be the FD set corresponding to $S_h$

\STATE Compute $C_{2opt}(\Sigma_h,I)$
\IF {$ |C_{2opt}(\Sigma_h,I)| \cdot \min\{|R|-1 , |\Sigma| \}  \leq \tau$}

\RETURN $\Sigma_h$

\ENDIF

\STATE remove $S_h$ from $PQ$
\FOR {each state $S_i$ that is a child of $S_h$}
\STATE let $\Sigma_i$ be the FD set corresponding to $S_i$
\STATE let $\mathcal{D}_s$ be the subset of difference sets in $\mathcal{D}$ that violate $\Sigma_i$
\STATE let $G_0$ be an empty graph
\STATE $minStates \leftarrow \texttt{getDescGoalStates}(S_i, S_i, G_0 , \mathcal{D}_s ,\tau)$
\STATE set $\underbar{gc}(S_i)$  to the minimum cost across all states in $minStates$, or $\infty$ if $minStates$ is empty

\IF {$\underbar{gc}(S_i)$ is not $\infty$}
\STATE insert $S_i$ into $PQ$
\ENDIF

\ENDFOR
\ENDWHILE

\RETURN $\phi$
\end{algorithmic}
\end{algorithm}

Algorithm~\ref{alg:astar} describes the search procedure. The goal of lines 1 and 12-16, along with the sub-procedure \texttt{getDescGoalStates}, is computing $\underbar{gc}(S)$. The reminder of Algorithm~\ref{alg:astar} follows the A* search algorithm: it initializes an open list, which is implemented as a priority queue called $PQ$, by inserting the root state $(\phi,\dots,\phi)$. In each iteration, the algorithm removes the state with the smallest value of $\underbar{gc}(S)$ from $PQ$ and checks whether it is a goal state. If so, the algorithm returns the corresponding FD set. Otherwise, the algorithm inserts the children of the removed state into $PQ$, after computing $\underbar{gc}(.)$ for each inserted state.

The two technical challenges of computing $\underbar{gc}(S)$ are the tightness of the bound $\underbar{gc}(S)$ (i.e., being close to the actual cost of the cheapest goal state descending from $S$), and having a small computational cost. In the following, we describe how we address these challenges.

Given a conflict graph $G$ of $I$ and $\Sigma$, each edge represents two tuples in $I$ that violate $\Sigma$. For any edge $(t_i,t_j)$ in $G$, we refer to the attributes that have different values in $t_i$ and $t_j$ as the \emph{difference set} of $(t_i,t_j)$. Difference sets have been introduced in the context of FD discovery (e.g., \cite{LopesPL00,WyssGR01}). For example, the difference sets for $(t_1,t_2)$, $(t_2,t_3)$, and $(t_3,t_4)$ in Figure~\ref{fig:conflict_graph} are $BD$, $AD$, and $BCD$, respectively. We denote by $\mathcal{D}$ the set of all difference sets for edges in $G$ (line 1 in Algorithm~\ref{alg:astar}).  The key idea that allows efficient computation of $\underbar{gc}(S)$ is that all edges (i.e., violations) in $G$ with the same difference set can be completely resolved by adding one attribute from the difference set to the LHS of each violated FD in $\Sigma$. For example, edges corresponding to difference set $BD$ in Figure~\ref{fig:conflict_graph} violate both $A \rightarrow B$ and $C \rightarrow D$, and to fix these violations, we need to add $D$ to the LHS of the first FD, and $B$ to the LHS of the second FD. Similarly, fixing violations corresponding to difference set $BCD$ can be done by adding $C$ or $D$ to the first FD (second FD is not violated).  Therefore, we partition the edges of the conflict graph $G$  based on their difference sets. In order to compute $\underbar{gc}(S)$, each group of edges corresponding to one difference set is considered atomically, rather than individually.

Let $\mathcal{D}_s$ be a \emph{subset} of difference sets that are still violated at the current state $S_i$ (line 13). Given a set of difference sets $\mathcal{D}_s$, the recursive procedure $\texttt{getDescGoalStates}(S, S_c, G_c, \mathcal{D}_c,\tau)$ (Algorithm~\ref{alg:minStates}) finds all minimal goal states descending from $S$ that resolve $\mathcal{D}_c$, taking into consideration the maximum number of allowed cell changes $\tau$. Therefore, $\underbar{gc}(S)$ can be assigned to the cheapest state returned by the procedure \texttt{getDescGoalStates}. Note that we use a subset of difference sets that are still violated  ($\mathcal{D}_s$), instead of using all violated difference sets, in order to efficiently compute $\underbar{gc}(S)$. The computed value of $\underbar{gc}(S)$ is clearly a lower bound on the cost of actual cheapest goal state descending from the current state $S$. To provide tight lower bounds, $\mathcal{D}_s$ is selected such that difference sets corresponding to large numbers of edges are favored. Additionally, we heuristically ensure that the difference sets in $\mathcal{D}_s$ have a small overlap.

We now describe Algorithm~\ref{alg:minStates}.  It recursively selects a difference set $d$ from the set of non-resolved difference sets $\mathcal{D}_c$. For each difference set $d$, we consider two alternatives: (1) excluding $d$ from being resolved, if threshold $\tau$ permits, and (2) resolving $d$ by extending the current state $S_c$. In the latter case, we consider all possible children of $S_c$ to resolve $d$. Once $S_c$ is extended to $S_c'$, we remove from $\mathcal{D}_c$ all the sets that are now resolved, resulting in $\mathcal{D}_c'$. Due to the monotonicity of the cost function, we can prune all the non-minimal states from the found set of states. That is, if state $S_1$ extends another state $S_2$ and both are goal states, we remove $S_1$. 

\begin{algorithm}
\small
\caption{$\texttt{getDescGoalStates}(S, S_c, G_c, \mathcal{D}_c, \tau)$}
\label{alg:minStates}
\begin{algorithmic}[1]
\REQUIRE $S$ : the state for which we compute $\underbar{gc}(.)$
\REQUIRE $S_c$ : the current state to be extended (equals $S$ at the first entry)
\REQUIRE $G_c$ : the current conflict graph for non-resolved difference sets (is empty at the first entry)
\REQUIRE $\mathcal{D}_c$ : the remaining difference sets to be resolved

\IF {$\mathcal{D}_c$ is empty }
\RETURN $\{S_c\}$
\ENDIF

\STATE $States \leftarrow \phi$

\STATE select a difference set $d$ from $\mathcal{D}_c$
\STATE let $G_c'$ be the graph whose edges are the union of edges corresponding to $d$ and edges of $G_c$
\STATE compute a 2-approximate minimum vertex cover of $G_c'$, denoted $C_{2opt}$
\IF { $ |C_{2opt}|\cdot \min\{|R|-1 , |\Sigma| \} < \tau$}

\STATE $\mathcal{D}_c' \leftarrow \mathcal{D}_c \setminus \{d\}$
\STATE $States \leftarrow States \cup \texttt{getDescGoalStates}(S,S_c, G_c', \mathcal{D}_c',\tau)$
\ENDIF

\FOR {each possible state $S_c'$ that extends $S_c$, is descendant of $S$, and resolves violations corresponding to $d$}
\STATE let $\mathcal{D}_c'$ be all difference sets in $\mathcal{D}_c$ that are still violating $\Sigma'_c$ that is corresponding to $S_c'$
\STATE $States \leftarrow States \cup \texttt{getDescGoalStates}(S, S_c', G_c, \mathcal{D}_c',\tau)$
\ENDFOR
\STATE remove any non-minimal states from $States$
\RETURN $States$
\end{algorithmic}
\end{algorithm}

In the following lemma, we prove that the computed value of $\underbar{gc}(S)$ is a lower bound on the cost of the cheapest goal descending from state $S$. The proof is in Appendix~\ref{sec:proof:lower_bound}.

\begin{lemma}
\label{lem:lower_bound}
For any state $S$,  $\underbar{gc}(S)$ is less than or equal to the cost of the cheapest goal state descendant of $S$.
\end{lemma}

%
%

Based on Lemma~\ref{lem:lower_bound}, and the correctness of the A* search algorithm \cite{PEARL89}, we conclude that the FD set generated by Algorithm~\ref{alg:astar} is part of a $P$-approximate $\tau$-constrained repair.

We now discuss the complexity of Algorithms~\ref{alg:astar} and \ref{alg:minStates}. Finding all difference sets in line 1 in  Algorithms~\ref{alg:astar} is performed in $O(|\Sigma| \cdot n + |\Sigma| \cdot |E| + |R| \cdot |E|)$, where $n$ denotes the number of tuples in $I$, and $E$ denotes the number of edges in the conflict graph of $I$ and $\Sigma$. Difference sets are obtained by building the conflict graph of $I$ and $\Sigma$, which costs $O(|\Sigma| \cdot n + |\Sigma| \cdot |E|)$ (more details are in Section~\ref{sec:datarep}), and then computing the difference set for all edges, which costs $O(|R| \cdot |E|)$.
In worst case, Algorithm~\ref{alg:astar}, which is based on A* search, will visit a number of states that is exponential in the depth of the cheapest goal state \cite{PEARL89}, which is less than $|\Sigma| \cdot (|R|-2)$. However, the number of states visited by an A* search algorithm is the minimum across all algorithms that traverse the same search tree and use the same heuristic for computing $\underbar{gc}(S)$. Also, we show in our experiments that the actual number of visited states is much smaller than the best-first search algorithm (Section~\ref{sec:expr}).

The worst-case complexity of Algorithm~\ref{alg:minStates} that finds $\underbar{gc}(S)$ is $O(|E|\cdot |R|^{|\Sigma| \cdot |\mathcal{D}_c|})$, where $|\mathcal{D}_c|$ is the number of difference sets passed to the algorithm.  This is due to recursively inspecting each difference set in $\mathcal{D}_c$ and, if not already resolved by the current state $S_c$,  appending one more attribute from the difference set to the LHS of each FD. At each step, approximate vertex graph cover might need to be computed, which can be performed in $O(|E|)$.

\section{Near-Optimal Data Modification}
\label{sec:datarep}

In this section, we derive a $P$-approximation of $\delta_{opt}(\Sigma',I)$, denoted $\delta_P(\Sigma',I)$, where $P=2 \cdot \min\{|R|-1,|\Sigma|\}$. We also give an algorithm that makes at most $\delta_P(\Sigma',I)$ cell changes in order to resolve all the inconsistencies with respect to the modified set of FDs computed in the previous section.

There are several data cleaning algorithms that obtain a data repair for a fixed set of FDs, such as \cite{BohannonFFR05,CongFGJM07,KolahiL09}. Most approaches do not provide any bounds on the number of cells that are changed during the repairing process. In \cite{KolahiL09}, the proposed algorithm provides an upper bound on the number of cell changes and it is proved to be near-minimum. The approximation factor depends on the set of FDs $\Sigma$, which is assumed to be fixed. Unfortunately, we need to deal with multiple FD sets, and the approximation factor described in \cite{KolahiL09} can grow arbitrarily while modifying the initial FD set. That is, the approximation factors for two possible repairs $\Sigma',\Sigma''$ in $\mathcal{S}(\Sigma)$ can be different. In this section, we provide a method to compute $\delta_P(\Sigma',I)$ such that the approximation factor is equal to $2 \cdot \min\{|R| -1 ,|\Sigma|\}$, which depends only on the number of attributes in $R$ and the number of FDs in $\Sigma$.

%
%
%

The output of our algorithm is a V-instance, which was first introduced in \cite{KolahiL09} to concisely represent multiple data instances (refer to Section~\ref{sec:notations} for more details). In the remainder of this paper, we refer to a V-instance as simply an instance.

The algorithm we propose in this section is a variant of the data cleaning algorithm proposed in \cite{BeskalesIG10}. The main difference is that we clean the data tuple-by-tuple instead of cell-by-cell. That is, we first identify a set of clean tuples that satisfy $\Sigma'$ such that the cardinality of the set is approximately maximum. We convert this problem to the problem of finding the minimum vertex cover, and we use a greedy algorithm with an approximation factor of 2. Then, we iteratively modify violating tuples as follows. For each violating tuple $t$, we iterate over attributes of $t$ in a random order, and we modify each attribute, if necessary, to ensure that the attributes processed so far are clean.

Given a set of FDs $\Sigma'$, the procedure \texttt{Repair\_Data} in Algorithm~\ref{alg:get_instance_repair} generates an instance $I'$ that satisfies $\Sigma'$. Initially, the algorithm constructs the conflict graph of $I$ and $\Sigma'$. Then, the algorithm obtains a 2-approximate minimum vertex cover of the obtained conflict graph, denoted $C_{2opt}(\Sigma',I)$, using a greedy approach described in \cite{GJ90} (for brevity, we refer to $C_{2opt}(\Sigma',I)$ as $C_{2opt}$ in this section). The clean instance $I'$ is initially set to $I$. The algorithm repeatedly removes a tuple $t$ from $C_{2opt}$, and it changes attributes of $t$ to ensure that, for every tuple $t' \in I' \setminus C_{2opt}$, $t$ and $t'$ do not violate $\Sigma'$ (lines 5-15). This is achieved by repeatedly picking an attribute of $t$ at random, and adding it to a set denoted $Fixed\_Attrs$ (line 9). After inserting an attribute $A$, we determine whether we can find an assignment to the attributes outside $Fixed\_Attrs$ such $(t,t')$ are not violating $\Sigma'$, for all $t' \in I' \setminus C_{2opt}$. We use Algorithm~\ref{alg:find_assignment} to find a valid assignment, if any, or to indicate that no valid assignment exists. Note that when $Fixed\_Attrs$ contains only one attribute (line 6), it is guaranteed that a valid assignment exists (line 7). If a valid assignment is found, we keep $t[A]$ unchanged. Otherwise, we change $t[A]$ to the value of attribute $A$ of the valid assignment found in the previous iteration (line 11). The algorithm proceeds until all tuples have been removed from $C_{2opt}$.  We return $I'$ upon termination.

\begin{algorithm}[h]
\small
\caption{\texttt{Repair\_Data($\Sigma'$,$I$)}}
\label{alg:get_instance_repair}
\begin{algorithmic}[1]

\STATE let $G$ be the conflict graph of $I$ and $\Sigma'$

\STATE obtain a 2-approximate minimum vertex cover of $G$, denoted $C_{2opt}$

\STATE $I' \leftarrow I$

\WHILE {$C_{2opt}$ is not empty}

    \STATE randomly pick a tuple $t$ from $C_{2opt}$

    \STATE $Fixed\_Attrs \leftarrow \{A\}$, where $A$ is a randomly picked attribute from $R$

    \STATE $t_c \leftarrow$ $\texttt{Find\_Assignment}(t,Fixed\_Attrs,I',\Sigma',C_{2opt})$

    \WHILE { $|Fixed\_Attrs| < |R|$}

        \STATE randomly pick an attribute $A$ from $R \setminus Fixed\_Atts$ and insert it into $Fixed\_Attrs$

        \IF { \texttt{Find\_Assignment}$(t, Fixed\_Attrs, I',\Sigma', C_{2opt}) = \phi$}

            \STATE $t[A]\leftarrow t_c[A]$
        \ELSE
            \STATE $t_c \leftarrow \texttt{Find\_Assignment}(t,Fixed\_Attrs,I',\Sigma',C_{2opt})$
        \ENDIF

    \ENDWHILE

    \STATE remove $t$ from $C_{2opt}$

\ENDWHILE

\RETURN $I'$

\end{algorithmic}
\end{algorithm}

\begin{algorithm}[t]
\small
\caption{\texttt{Find\_Assignment}$(t,Fixed\_Attrs,I',\Sigma',C_{2opt})$}
\label{alg:find_assignment}
\begin{algorithmic}[1]

\STATE construct a tuple $t_c$ such that $t_c[A] =t[A]$ if $A\in Fixed\_Attrs$, and $t_c[A] = v_i^A$ if $A \not \in Fixed\_Attrs$, where $v_i^A$ is a new variable

\WHILE {$\exists t' \in I' \setminus C_{2opt}$ such that for some FD $X \rightarrow A \in \Sigma'$, $t_c[X]=t'[X] \wedge t_c[A]\neq t'[A]$}

\IF {$A \in Fixed\_Attrs$}

\RETURN $\phi$

\ELSE

\STATE $t_c[A] \leftarrow t'[A]$

\STATE add $A$ to $Fixed\_Attrs$
\ENDIF

\ENDWHILE

\RETURN $t_c$

\end{algorithmic}
\end{algorithm}

\begin{figure}[h!]
  \center
  \includegraphics[width=2.75in]{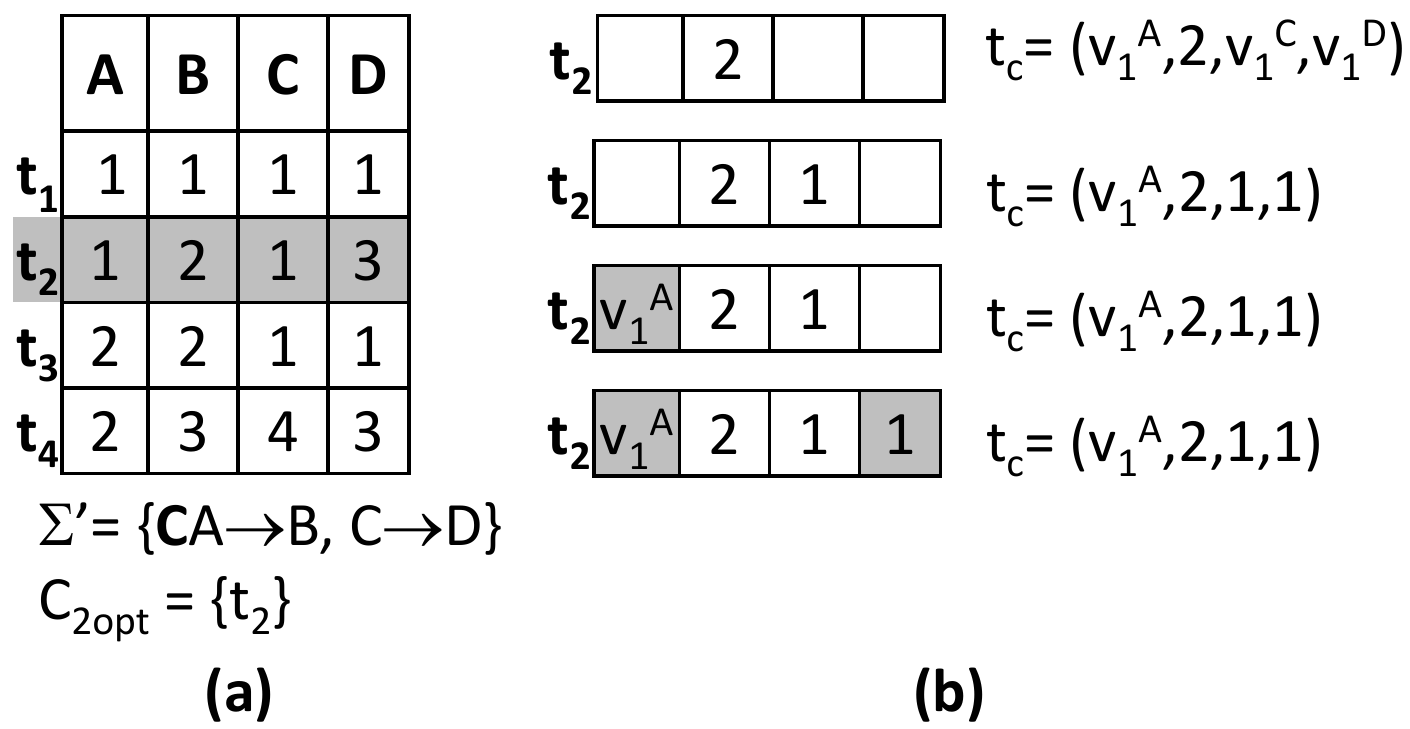}\\
  \caption{An example of repairing data: (a) initial value of $I'$, $\Sigma'$ and $C_{2opt}$ (b) steps of fixing the tuple $t_2$}\label{fig:fix_tuple}
\end{figure}

Algorithm~\ref{alg:find_assignment} searches for an assignment to attributes of a tuple $t$ that are not in $Fixed\_Attrs$ such that every pair $(t,t')$ satisfies $\Sigma'$ for all $t' \in I' \setminus C_{2opt}$. An initial assignment $t_c$ is created by setting attributes that are in $Fixed\_Attrs$ to be equal to $t$, and setting attributes that are not in $Fixed\_Attrs$ to new variables. The algorithm repeatedly selects a tuple $t' \in I' \setminus C_{2opt}$ such that $(t,t')$  violates an FD $X\rightarrow A \in \Sigma'$. If attribute $A$ belongs to $Fixed\_Attrs$, the algorithm returns $\phi$, indicating that no valid assignment is available. Otherwise, the algorithm sets $t[A]$ to be equal to $t'[A]$, and adds $A$ to $Fixed\_Attrs$. When no other violations could be found, the algorithm returns the assignment $t_c$.

In Figure~\ref{fig:fix_tuple}, we show an example of generating a data repair for $\Sigma'= \{CA \rightarrow B, C \rightarrow D\}$, given the instance $I$ shown in Figure~\ref{fig:fix_tuple}(a). After adding the first attribute $B$ to $Fixed\_Attrs$, the current valid assignment, denoted $t_c$, is equal to $(v_1^A,2, v_1^C, v_1^D)$. When inserting $C$ to $Fixed\_Attrs$, there is no need to change the value of $C$ because we can find a valid assignment to the remaining attributes, which is $(v_1^A, 2,1,1)$. After inserting $A$ to $Fixed\_Attrs$, no valid assignment is found, and thus we set $t[A]$ to the value of attribute $A$ of the previous valid assignment $t_c$. Similarly, we set $t[D]$ to $t_c[D]$ after inserting $D$ into $Fixed\_Attrs$. The resulting instance satisfies $\Sigma'$.
The following lemma proves the soundness and completeness of Algorithm~\ref{alg:find_assignment}. The proof is in Appendix~\ref{sec:proof:find_assignment}.

\begin{lemma}
\label{lem:find_assignment}
Algorithm~\ref{alg:find_assignment} is both sound (i.e., the obtained assignments are valid) and complete (it will return an assignment if a valid assignment exists).
\end{lemma}

The following theorem proves the $P$-optimality of Algorithm~\ref{alg:get_instance_repair}. The proof is in Appendix~\ref{sec:proof:P}.

\begin{theorem}
\label{thm:P}

For a given instance $I$ and a set of FDs $\Sigma' \in \mathcal{S}(\Sigma)$, Algorithm \texttt{Repair\_Data($\Sigma',I$)} obtains an instance $I' \models \Sigma'$ such that the number of changed cells in $I'$ is at most $|C_{2opt}(\Sigma',I)| \cdot \min\{|R|-1,|\Sigma|\}$, and it is $2 \cdot \min\{|R|-1,|\Sigma|\}$-approximate minimum.
\end{theorem}

%

We now describe the worst-case complexity of Algorithms~\ref{alg:get_instance_repair} and \ref{alg:find_assignment}. Algorithm~\ref{alg:find_assignment} has a complexity of $O(|R|+|\Sigma'|)$ because constructing $t_c$ in line 1 costs $O(|R|)$, and the loop in lines 2-9 iterates at most $|\Sigma'|$ times. The reason is that, for each FD $X\rightarrow A \in \Sigma'$, there is at most one tuple in $I' \setminus C_{2opt}$ satisfying the condition in line 2 (otherwise, tuples in $I' \setminus C_{2opt}$ would be violating $X \rightarrow A$).

Constructing the conflict graph in line 1 in Algorithm~\ref{alg:get_instance_repair} takes $O(|\Sigma'| \cdot n + |\Sigma'| \cdot |E|)$, where $|\Sigma'|$ is the number of FDs in $\Sigma'$, $n$ is the number of tuples in $I$ and $E$ is the set of edges in the resulting conflict graph. This step is performed by partitioning tuples in $I$ based on LHS attributes of each FD in $\Sigma'$ using a hashing function, and constructing sub-partitions within each partition based on right-hand-side attributes of each FD. Edges of the conflict graph are generated by emitting pairs of tuples that belong to the same partition and different sub-partitions. The approximate vertex cover is computed in $O(|E|)$ \cite{GJ90}. The loop in lines 4-17 iterates a number of times equal to the size of the vertex cover, which is $O(n)$. Each iteration costs $O(|R| \cdot (|R| + |\Sigma'|))$.  To sum up, the complexity of finding a clean instance $I'$ is $O(|\Sigma'| \cdot |E| + |R|^2 \cdot n + |R| \cdot |\Sigma'| \cdot n)$. Assuming that $|R|$ and $|\Sigma'|$ are much smaller than $n$, the complexity is reduced to $O(|E| + n)$.

%

\section{Computing Multiple Repairs}
\label{sec:range}

So far, we discussed how to modify the data and the FDs for a given value of $\tau$.
One way to obtain a small sample of possible repairs is to execute Algorithm~\ref{alg:repair} multiple times with randomly chosen values of $\tau$. This can be easily parallelized, but may be inefficient for two reasons. First, multiple values of $\tau$ could result in the same repair, making some executions of the algorithm redundant. Second, different invocations of Algorithm~\ref{alg:astar} are expected to visit the same states, so we should be able to re-use previous computations. To overcome these drawbacks, we develop an algorithm (Algorithm~\ref{alg:range}) that generates minimal FD modifications corresponding to a range of $\tau$ values. We then use Algorithm~\ref{alg:get_instance_repair} to find the corresponding data modifications.

\begin{algorithm}[t]
\small
\caption{\texttt{Find\_Repairs\_FDs($\Sigma,I,\tau_l,\tau_u$)}}
\label{alg:range}

\begin{algorithmic}[1]

\STATE $PQ \leftarrow \{ (\phi,\dots,\phi) \}$
\STATE $\tau \leftarrow \tau_u$
\STATE $FD\_Repairs \leftarrow \phi$

\WHILE {$PQ$ is not empty and $\tau \geq \tau_l$}

\STATE Pick the state $S_h$ with the smallest value of $\underbar{gc}(.)$ from $PQ$

\STATE Let $\Sigma_h$ be the FD set corresponding to $S_h$

\STATE Compute $C_{2opt}(\Sigma_h,I)$
\IF {$|C_{2opt}(\Sigma_h,I)| \cdot \min\{|R|-1,|\Sigma|\} \leq \tau$}

\STATE Add $\Sigma_h$ to $FD\_Repairs$

\STATE $\tau \leftarrow |C_{2opt}(\Sigma_h,I)| \cdot \min\{|R|-1,|\Sigma|\} - 1$

\STATE For each state $S_i \in PQ$, recompute $\underbar{gc}(S_i)$ using the new value of $\tau$

\ENDIF

\STATE Remove $S_h$ from $PQ$
\FOR {each state $S_i$ that is a child of $S_h$}
\STATE Compute $\underbar{gc}(S_i)$ (similar to Algorithm~\ref{alg:astar})
\STATE Insert $S_i$ into $PQ$
\ENDFOR
\ENDWHILE

\RETURN $FD\_Repairs$
\end{algorithmic}
\end{algorithm}

Algorithm~\ref{alg:range} generates FD modifications corresponding to the relative trust range $\tau \in [\tau_l,\tau_u]$. Initially, $\tau=\tau_u$. We proceed by visiting states in order of $\underbar{gc}(.)$, and expanding $PQ$ by inserting new states. Once a goal state is found, the corresponding FD modification $\Sigma_h$ is added to the set of possible repairs. The set $\Sigma_h$ corresponds to the trust range $[\delta_P(\Sigma_h,I),\tau]$. Therefore, we set the new value of $\tau$ to $\delta_P(\Sigma_h,I) -1$ in order to discover a new repair. Because $\underbar{gc}(.)$ depends on $\tau$, we recompute $\underbar{gc}(.)$ for all states in $PQ$. Note that states that have been previously removed from $PQ$ because they were not goal states (line 13) cannot be goal states with respect to the new value of $\tau$.  The reason is that if a state is not a goal state for $\tau=x$, it cannot be a goal state for $\tau < x$ (refer to line 8). The algorithm terminates when $PQ$ is empty, or when $\tau < \tau_l$.  Finally, we take all the FD modifications we found (or a sample of them if we have found too many), and we generate the corresponding data modifications.

\section{Experimental Evaluation}
\label{sec:expr}

In this section, we study the relationship between the quality of repairs and the relative trust determined by $\tau$, and we compare our approach to the technique introduced in \cite{ChiangM11}. Also, we show the efficiency of our repair generating algorithms.

\subsection{Setup}

All experiments were conducted on a SunFire X4100 server with a Quad-Core 2.2GHz processor, and 8GB of RAM. All computations are executed in memory. Repairing algorithms are executed as single-threaded processes, and we limit memory usage to 1.5GB. We use a real data set, namely the Census-Income data set\footnote{http://archive.ics.uci.edu/ml/datasets/Census-Income+(KDD)}, which is part of the UC Irvine Machine Learning Repository. Census-Income consists of 300k tuples and 40 attributes (we only use 34 attributes in our experiments). To perform experiments on smaller data sizes, we randomly pick a sample of tuples.

We tested two variants of Algorithm~\texttt{Repair\_Data\_FDs}: \texttt{$A^*$-Repair} which uses the A*-based search algorithm described in Section~\ref{sec:astar}, and \texttt{Best-First-Repair} which uses a best-first search to obtain FD repairs, as we described in Section~\ref{sec:cheapestFDs}. Both variants use Algorithm~\ref{alg:get_instance_repair} to obtain the corresponding data repair. We use the number of distinct values  to measure the weights of sets of attributes appended to LHS's of FDs (i.e., $w(Y) = F_{count(Y)} \Pi_Y(I)$ ). In our experiments, we adjust the relative threshold $\tau_r$, rather than the absolute threshold $\tau$. 
We also implemented the repairing algorithm introduced in \cite{ChiangM11}, which uses a unified cost model to quantify the goodness of each data-FD repair and obtains the repair with the (heuristically) minimum cost.

In order to assess the quality of the generated repairs, we first use an FD discovery algorithm to find all the minimal FDs with a relatively small number of attributes in the LHS (less than 6). In each experiment, we randomly select a number of FDs from the discovered list of FDs. We denote by $I_c$ and $\Sigma_c$ the clean database instance and the FDs, respectively. The data instance $I_c$ is perturbed by changing the value of some cells such that each cell change results in a violation of an FD. Specifically, we inject two types of violations as follows.

\begin{itemize}
\item Right-hand-side violation: We first search for two tuples $t_i,t_j$ that agree on $XA$ for some FD $X \rightarrow A \in \Sigma$. Then, we modify $t_i[A]$ to be different from $t_j[A]$.
\item Left-hand-side violation: We search for two tuples $t_i,t_j$ such that for some FD $X \rightarrow A$, $t_i[X \setminus \{B\}] = t_j[X \setminus \{B\}]$, $t_i[B] \neq t_j[B]$ and $t_i[A] \neq t_j[A]$, where $B \in X$. We introduce a violation by setting $t_i[B]$ to $t_j[B]$.
\end{itemize}

We refer to the resulting instance as $I_d$. In our approach, we concentrate on one method of fixing FDs, which is appending one or more attributes to LHS's of FDs. Therefore, we perform FDs perturbation by randomly removing a number of attributes from their LHS's. The perturbed set of FDs is denoted $\Sigma_d$. The cleaning algorithm is applied to $(\Sigma_d,I_d)$, and the resulting repair is denoted $(\Sigma_r,I_r)$. The parameters that control the perturbation of data and FDs are (1) Data Error Rate, which is the fraction of cells that are modified, and (2) FD Error Rate, which is the fraction of LHS attributes that were removed. We use the following metrics to measure the quality of the modified data and FDs.

\begin{itemize}
\item Data precision: the ratio of the number of correctly modified cells to the total number of cells modified by the repair algorithm. A modification of a cell $t[A]$ is considered correct if the values of $t[A]$ in $I_c$ and $I_d$ are different, and either $t[A]$ in $I_r$ is equal to $t[A]$ in $I_c$, or $t[A]$ is a variable in $I_r$.

\item Data recall: the ratio of the number of correctly modified cells to the total number of erroneous cells (i.e., cells with different values in $I_d$ and $I_c$).

\item FD precision: the ratio of the number of attributes correctly appended to LHS's of FDs in $\Sigma_d$ to the total number of appended attributes.

\item FD recall: the ratio of the number of attributes correctly appended to LHS's of FDs in $\Sigma_d$ to the total number of attributes removed from $\Sigma_c$ while constructing $\Sigma_d$.

\end{itemize}

In order to measure the overall quality of a repair $(\Sigma_r,I_r)$, we compute the harmonic averages of precision and recall for both data and FDs (also called F-scores). Then, we compute the average F-score for data and FDs, which we refer to as the combined F-score.

\subsection{Impact of Relative Trust on Repair Quality}

\begin{figure}
  \center
  \includegraphics[width=3.3in]{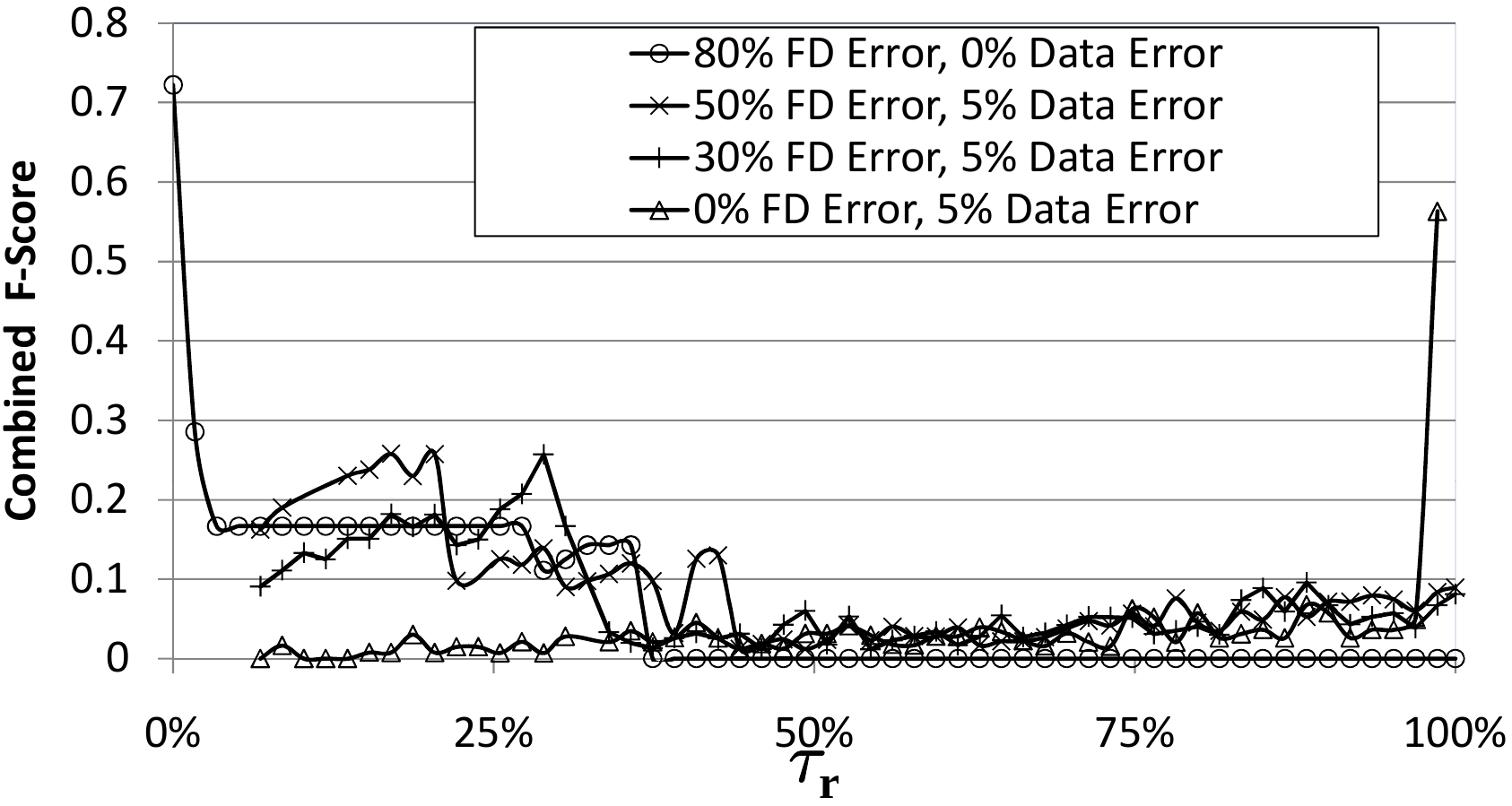}\\
  \caption{{Repair quality at multiple error rates}}\label{fig:expr_qual}
\end{figure}

In this experiment, we measure the combined F-score at various error rates. We use 5000 tuples from the Census-Income data set to represent the clean instance $I_c$, and we use an FD with 6 LHS attributes to represent $\Sigma_c$.  Figure~\ref{fig:expr_qual} shows the combined F-score for various data sets, for multiple values of $\tau_r$. When only FDs perturbation is performed, we notice that the peak quality occurs at $\tau_r = 0 \%$ (i.e., when no changes to data are allowed). At FD error rate of $50\%$, we notice that the peak quality occurs at $\tau_r = 17\%$. At 30\% FD error rate and 5\% data error rate, the peak quality occurs at higher value ($\tau_r=28.9\%$). Finally, when only data perturbation is performed, the peak quality occurs at $\tau=100\%$ (i.e., the algorithm can freely change the data, while obtaining the cheapest FD repair, which is the original FD). 

\begin{figure}
  \center
  \includegraphics[width=3.3in]{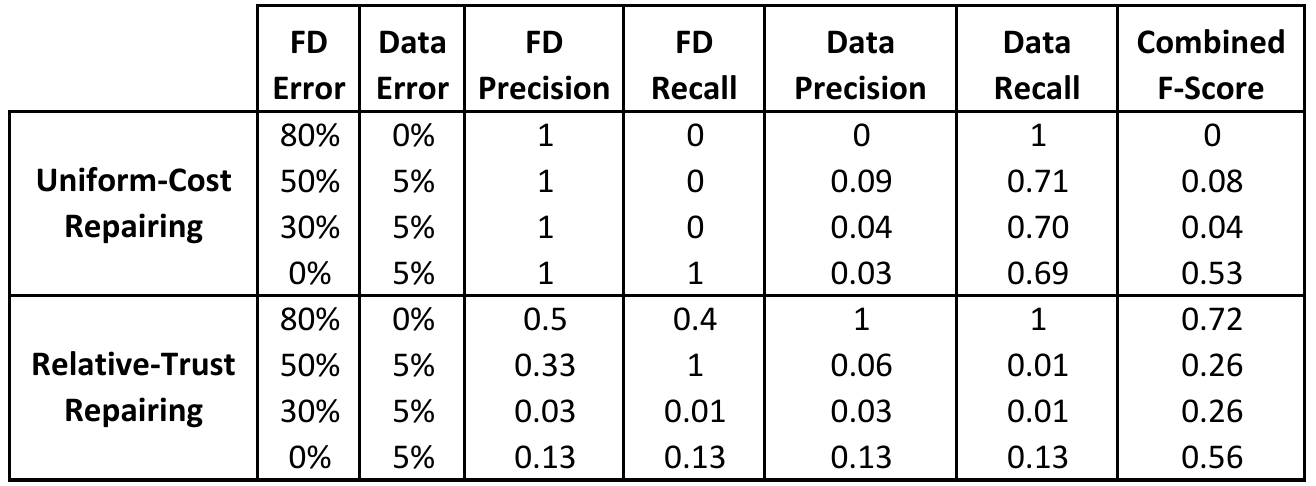}\\
  \caption{The maximum quality achievable by our algorithm and the algorithm in \protect\cite{ChiangM11}}\label{fig:expr_fei}
\end{figure}

In Figure~\ref{fig:expr_fei}, we compare the quality of the repairs generated by our algorithm, denoted Relative-Trust Repairing, to the quality of repairs generated by the repairing algorithm in \cite{ChiangM11}, denoted Uniform-Cost Repairing. For both algorithms, we tested multiple parameter settings and we reported the quality metrics of the repair with the highest combined F-score. For example, for FD error of 50\% and data error of 5\% our algorithm achieved the maximum combined F-score of 0.26 at $\tau = 17\%$. For all data sets, we noticed that the algorithm in \cite{ChiangM11} did not choose to modify the FD using any parameter settings, resulting in FD precision of 1 and recall of 0 for the first three data sets, and recall of 1 for the fourth data set. Because our algorithm is aware of the different levels of relative trusts, we were able to achieve higher quality scores when choosing the appropriate value of $\tau$. This is clear in the first data set with FD error of 80\% and data error of 0\%. Insisting on modifying the data, not the FD, resulted in FD recall of 0, and data precision of 0. On the other hand, when setting $\tau$ to $0\%$, our algorithm kept the data unmodified, resulting in perfect data precision and recall, and changed the FD, resulting in FD precision of 0.5 and FD recall of 0.4.

Note that, in general, the precision and recall for data repairs is relatively low due to the high uncertainty about the right cells to modify. For example, given an FD $A \rightarrow B$, and two violating tuples $t_1$ and $t_2$, we have four cells that can be changed in order to repair the violation: $t_1[A]$, $t_1[B]$, $t_2[A]$, and $t_2[B]$. This can be reduced by considering additional information such as the user trust in various attributes and tuples (e.g., \cite{BohannonFFR05,ChiangM11,KolahiL09}). Using this kind of information is not considered in our work.

\subsection{Performance Results}

\subsubsection{Scalability with the Number of Tuples}

\begin{figure}[t]
  \center
  \includegraphics[width=3.4in]{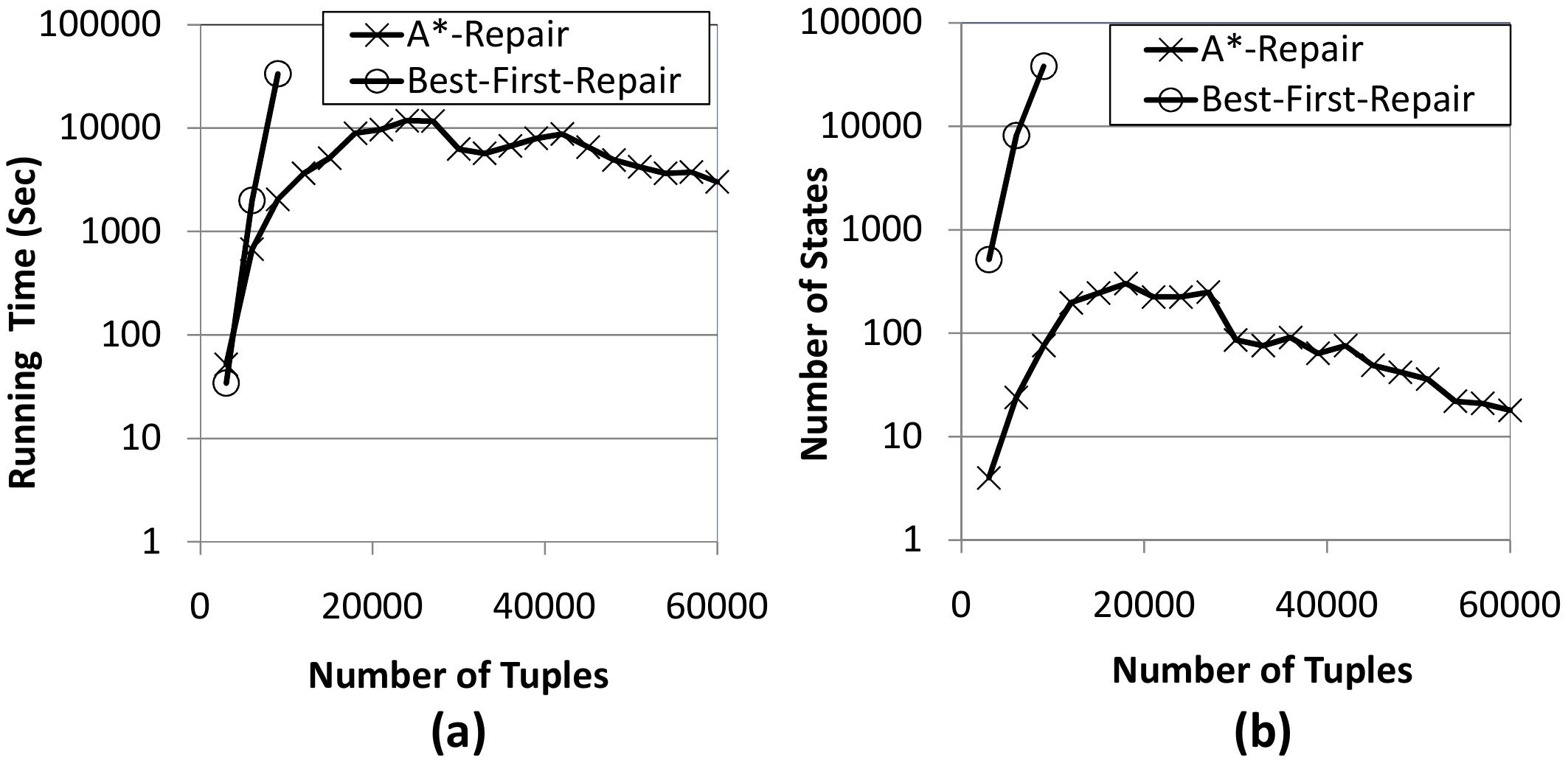}\\
  \caption{{Performance at various instance sizes}}\label{fig:expr_perf_tuples}
\end{figure}

In this experiment, we show the scalability of our algorithms with respect to the number of tuples. We use two FDs, and we set $\tau_r$ to $1\%$. Figures~\ref{fig:expr_perf_tuples}(a) and \ref{fig:expr_perf_tuples}(b) show the running time, and the number of visited states, respectively, against the number of tuples. When increasing the number of tuples in the range $[0,20000]$, the number of unique difference sets increases, while the average frequency of difference sets remains relatively small, compared to $\tau$. It follows that the computed lower bounds $\underbar{gc}(S)$ are very loose because most difference sets considered by Algorithm~\ref{alg:minStates} can be left unresolved (i.e., the condition in line 8 is true). Thus, the search algorithm needs to visit more states, as we show in Figure~\ref{fig:expr_perf_tuples}(b). 

When the number of tuples increases beyond 20000, we notice in Figure~\ref{fig:expr_perf_tuples} that the running time, as well as the number of visited states, decreases. The reason is that, in the state searching algorithm, the number of distinct difference sets stabilizes after reaching a certain number of tuples, and the frequencies of individual difference sets start increasing. It follows that most difference sets can no longer remain unresolved, and tighter lower bounds $\underbar{gc}(S)$ are reported, which leads to decreasing the number of visited states (Figure~\ref{fig:expr_perf_tuples}(b)).

Algorithm~\texttt{Best-First-Repair} does not depend on cost estimation, and thus, the execution time rapidly grows with the number of tuples in the entire range $[0,60000]$.

%

\subsubsection{Scalability with the Number of Attributes}

\begin{figure}
\center
\begin{minipage}{0.45\columnwidth}
\center
 \includegraphics[width=1\linewidth,clip]{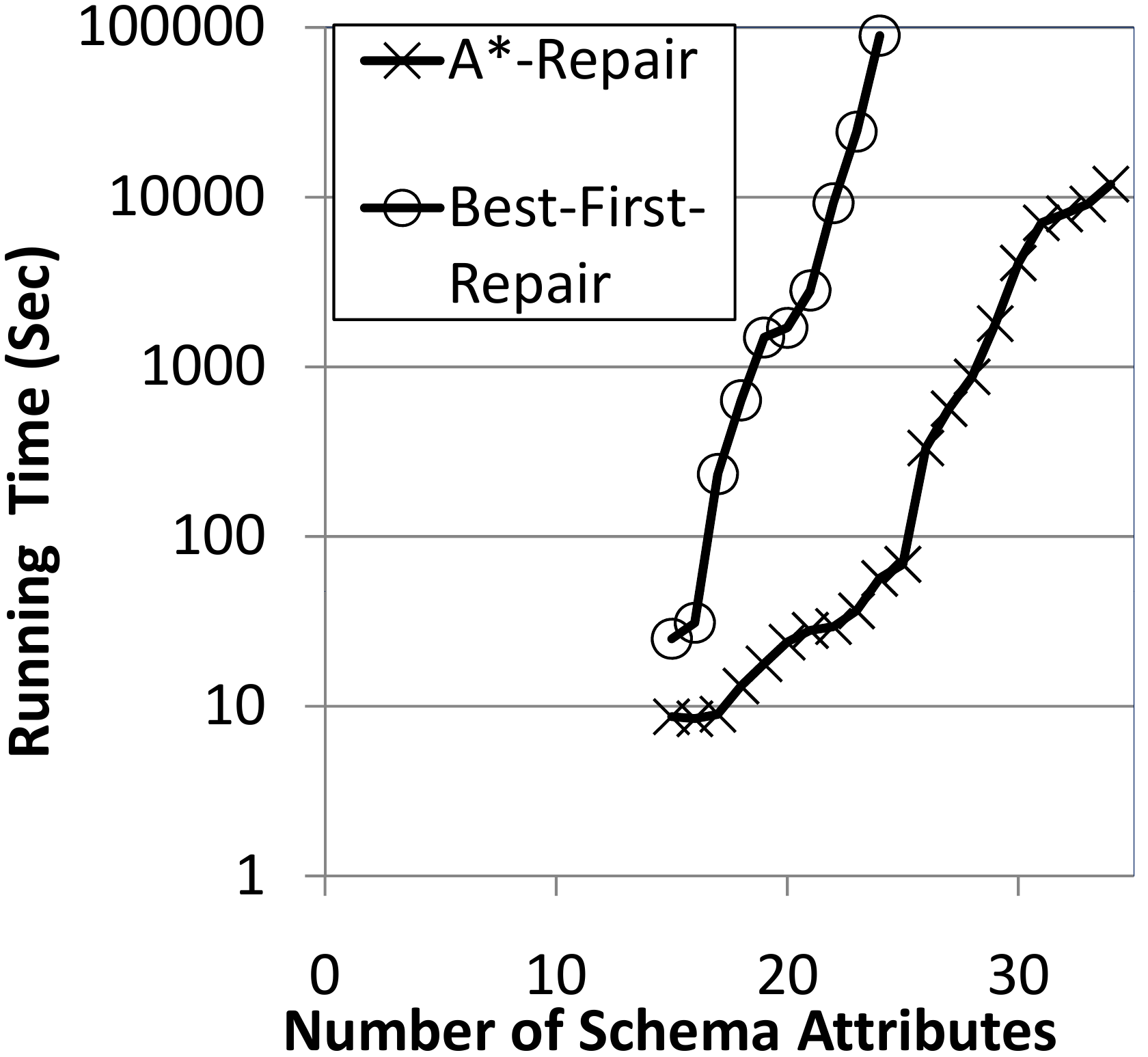}
  \caption{{Salability with number of attributes}}\label{fig:expr_perf_attrs}
\end{minipage}
\hspace{0.05in}
\begin{minipage}{0.45\columnwidth}
\centering
 \includegraphics[width=1\linewidth,clip]{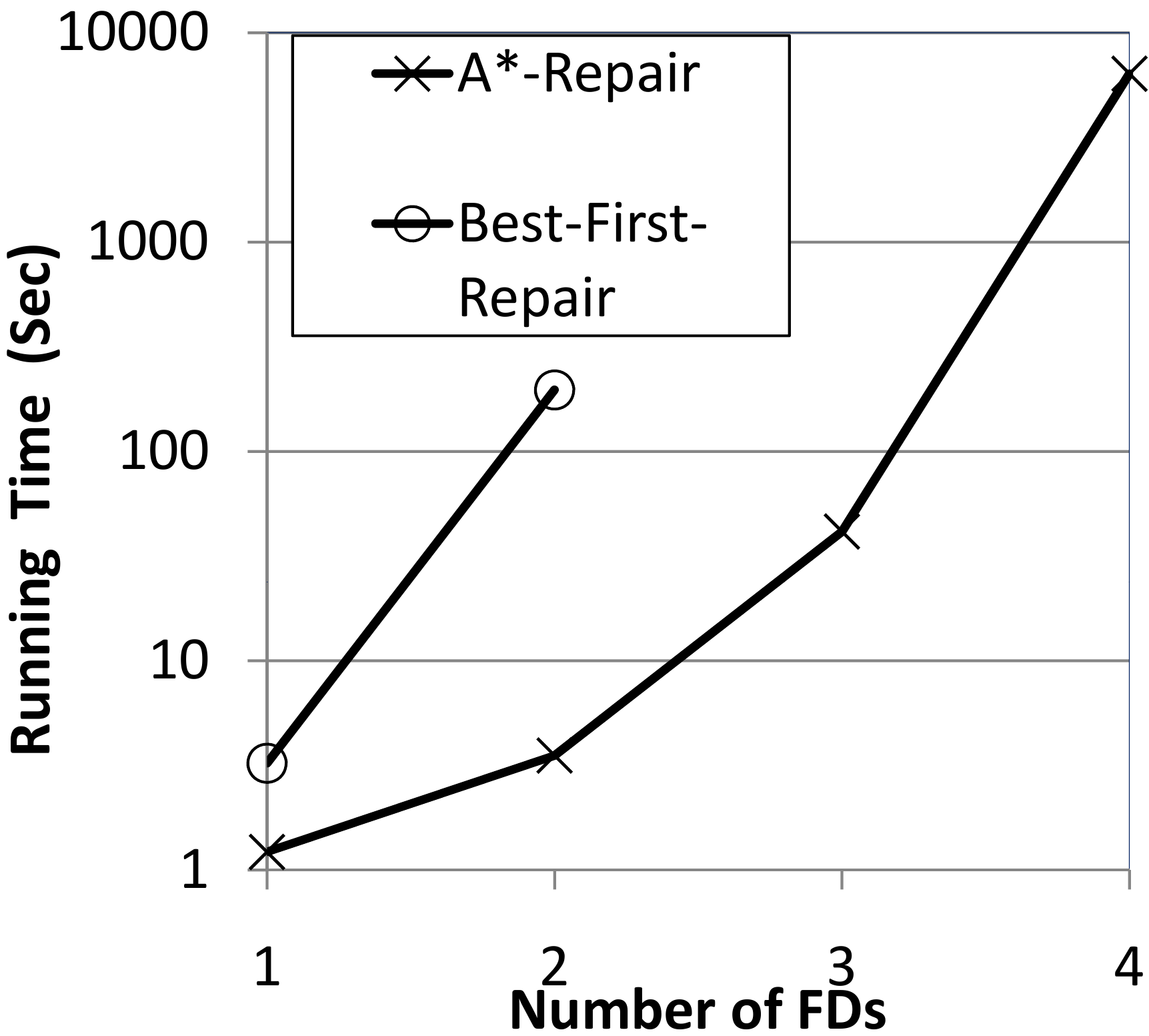}
  \caption{{Salability with number of FDs}}\label{fig:expr_perf_fds}
\end{minipage}
\end{figure}


Figure~\ref{fig:expr_perf_attrs} depicts the scalability of our approach with respect to the number of attributes. In this experiment, we used two FDs and 24000 tuples, and we set $\tau_r$ to $1\%$. We changed the number of attributes by excluding some number of attributes from the input relation. The running time increases with the number of attributes mainly because the size of state space increases exponentially with the number of attributes. 

\subsubsection{Scalability with the Number of FDs}

Figure~\ref{fig:expr_perf_fds} depicts the scalability of our approach with respect to the number of FDs. In this experiment, we used 10000 tuples, and we set $\tau_r$ to $1\%$. We use a single FD, and we replicate this FD multiple times to simulate larger sizes of $\Sigma$. The size of state space grows exponentially with the number of FDs. Thus, the searching algorithm visits more states, which increases the overall running time for both approaches: \texttt{$A^*$-Repair} and \texttt{Best-First-Repair}. Note that the  algorithm \texttt{Best-First-Repair} did not terminate in 24 hours when the number of FDs is greater than 2.

\subsubsection{Effect of the Relative Trust Parameter $\tau$}

\begin{figure}
  \center
  \includegraphics[width=3.4in]{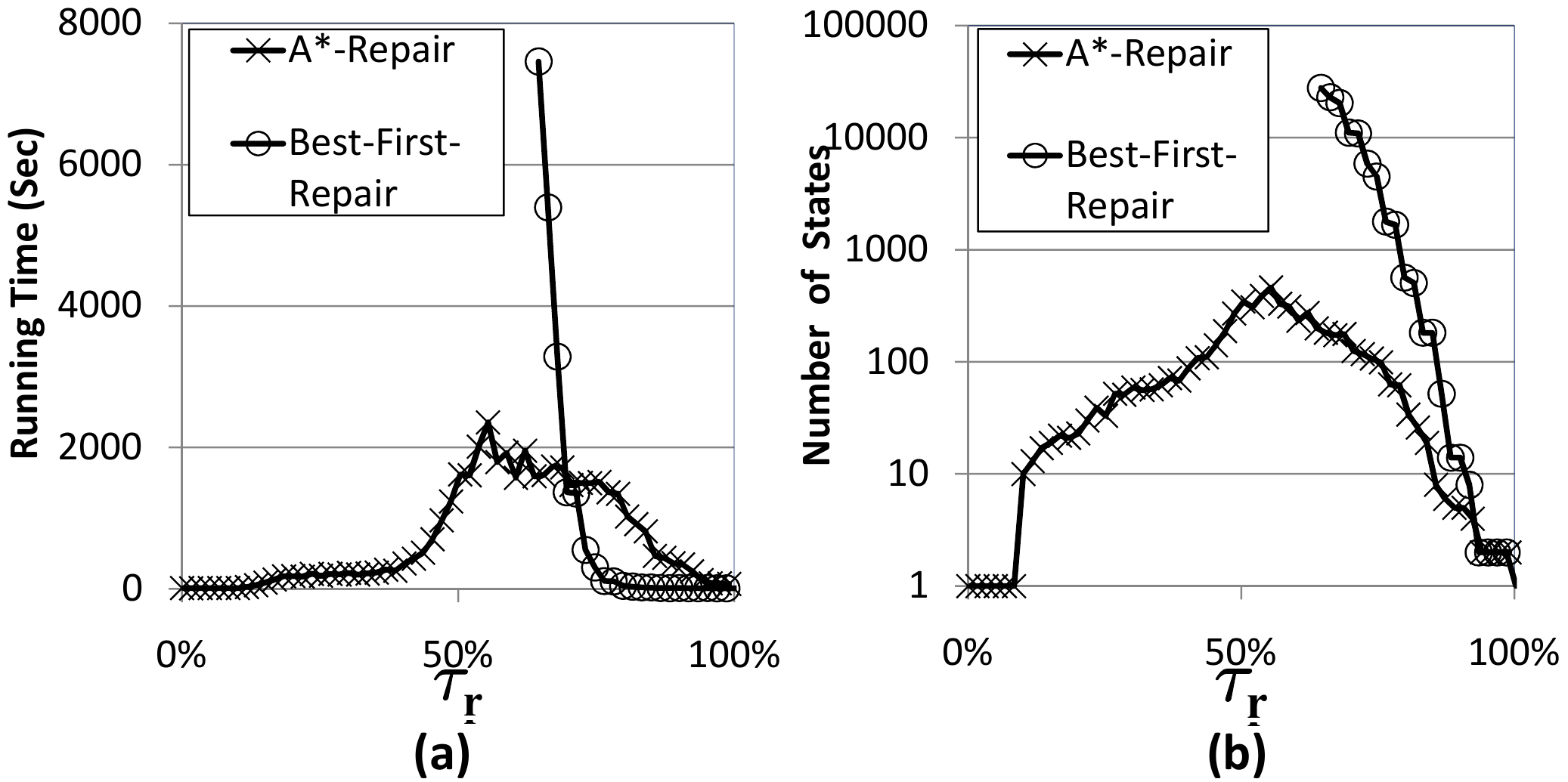}\\
  \caption{Effect of $\tau$ on (a) running time (b) visited states} \label{fig:expr_perf_tau}
   \vspace{-.15in}
\end{figure}

Figures~\ref{fig:expr_perf_tau}(a) and \ref{fig:expr_perf_tau}(b) show the running time and the number of visited states, respectively, for various values of $\tau_r$. In this experiment, we fix the number of tuples to be 5000, and we use $\Sigma_d$ with one FD. The number of appended attributes ranges from 9 at $\tau_r=10\%$ to 1 at $\tau_r=99\%$. No repair could be found for $\tau_r$ less than $10\%$.
We notice that at small values of $\tau$, Algorithm \texttt{$A^*$-Repair} is orders of magnitude faster than Algorithm \texttt{Best-First-Repair}. This is due to the effectiveness (i.e., tightness) of the cost estimation implemented in Algorithm \texttt{$A^*$-Repair}.  The lack of such estimation causes Algorithm \texttt{Best-First-Repair} to visit many more states.

As the value of $\tau_r$ increases up to $55\%$, we observe that Algorithm \texttt{$A^*$-Repair} becomes slower. The reason is that larger values of $\tau_r$ decreases the tightness of computed bounds $\underbar{gc}(S)$. As $\tau_r$ increases beyond $55\%$, we notice an improvement in the running time as we only need to add very small number of attributes to reach a goal state. 

\subsubsection{Generating Multiple Repairs}

\begin{figure}
  \center
  \includegraphics[width=1.8in]{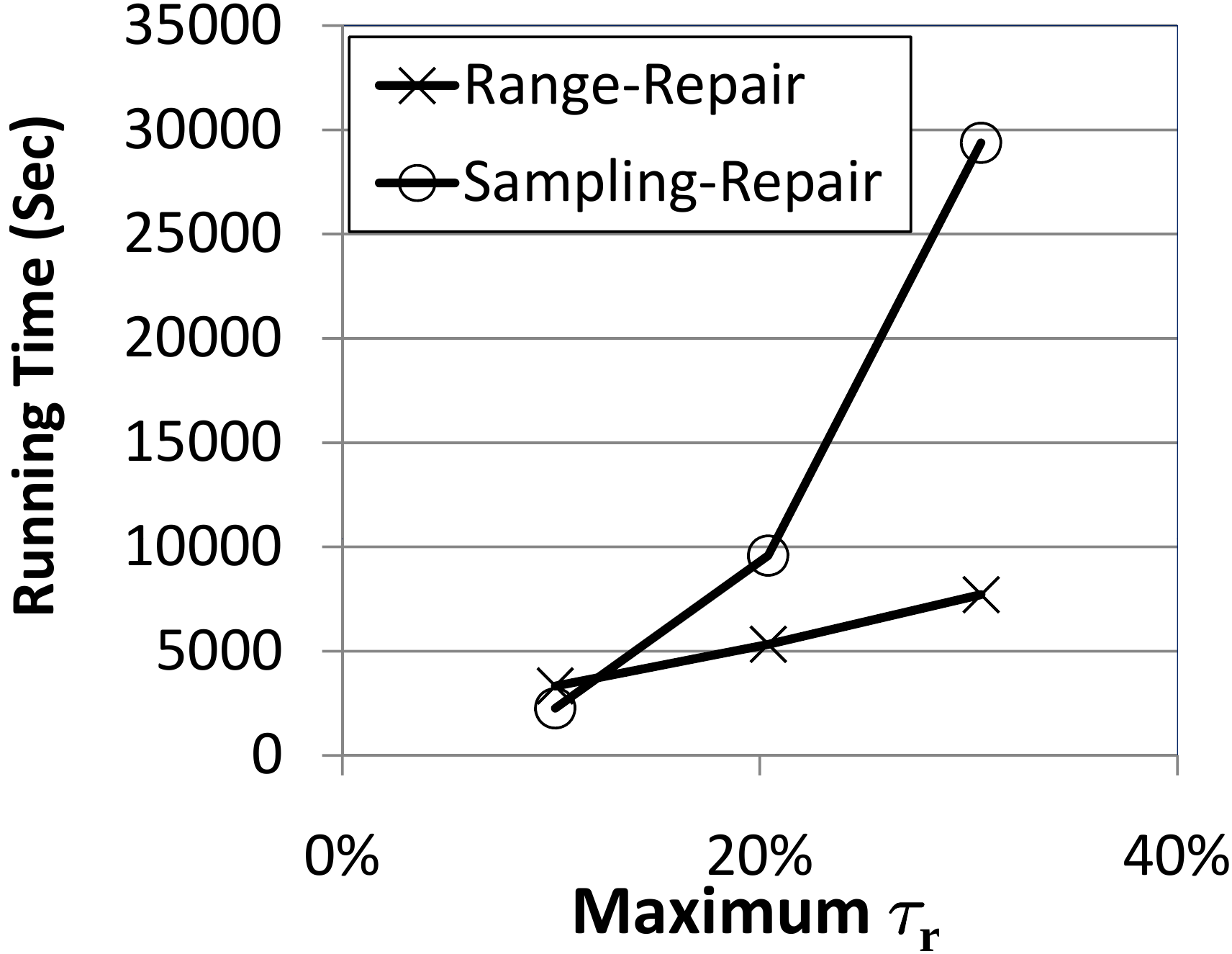}\\
    \vspace{-.15in}
  \caption{Performance under uncertain relative trust}\label{fig:expr_range}
   \vspace{-.2in}
\end{figure}

In this experiment, we assess the efficiency of two approaches that generate possible repairs for a given range of $\tau_r$. In the first approach, denoted \texttt{Range-Repair}, we execute Algorithm~\ref{alg:range}, and we invoke the data repair algorithm (Algorithm~\ref{alg:get_instance_repair}) for each obtained FD repair. In the second approach, denoted \texttt{Sampling-Repair}, we invoke the algorithm \texttt{$A^*$-Repair} at a sample of possible values of $\tau_r$. In this experiment, we used 5000 tuples, and one FD. We set the minimum value of $\tau_r$ to 0, and we varied the upper bound of $\tau$ in the range  $[10\%,30\%]$, which is represented by the X-axis in Figure~\ref{fig:expr_range}.  For the sampling approach, we started by $\tau_r =0\%$, and we increased $\tau_r$ in steps of $1.7\%$ (which is equal to $13$ in this experiment) until we reach the maximum value of $\tau_r$.  Figure~\ref{fig:expr_range} shows the running time for both approaches. We observe that \texttt{Range-Repair} outperforms the sampling approach, especially at wide ranges of $\tau_r$. For example, for the range $[0,30\%]$, \texttt{Range-Repair} is 3.8 times faster than \texttt{Sampling-Repair}.

\section{Related Work}
\label{sec:related}

The closest work to ours is \cite{ChiangM11}, which proposed a technique to obtain a single repair, $(\Sigma',I')$, of the FDs and the data, respectively, for a given input $(\Sigma,I)$. A unified cost model was proposed to measure the distance between a repair $(\Sigma',I')$ and the inputs $(\Sigma,I)$. An approximate algorithm was presented that obtains a repair with the minimum cost.  There are many differences between our work and \cite{ChiangM11} including: 1) we incorporate the notion of relative trust in the data cleaning process and produce multiple suggested repairs corresponding to various levels of relative trust; 2) \cite{ChiangM11} does not give any minimality guarantees for the generated repairs; 3) the algorithm proposed in \cite{ChiangM11} searches a constrained repair space that only considers adding single attributes to LHS's of FDs in $\Sigma$, while we explore a larger repair space that considers appending any subset of $R$ to the LHS of each FD.


The idea of modifying a supplied set of FDs to better fit the data was also discussed in \cite{GolabKKS10}.  The goal of that work was to generate a small set of Conditional Functional Dependencies (CFDs) by modifying the embedded FD.  Modifying the data and relative trust were not discussed.

The problem of cleaning the data in order to satisfy a fixed set of FDs has been studied in, e.g., \cite{BeskalesIG10,BohannonFFR05,KolahiL09,CongFGJM07}.  In our context, these solutions may be classified as having a fixed threshold $\tau_r$ of $100\%$.  Part of our work is inspired by the algorithm proposed in \cite{BeskalesIG10} in the sense that we incrementally modify the data until there are no inconsistencies left. However, we modify individual tuples instead of attribute values. Also, unlike the approach in \cite{BeskalesIG10}, we provide an upper bound on the number of changes.

Another related problem is discovering which FDs hold (approximately or exactly) on a fixed database instance; see, e.g., \cite{LopesPL00,KramerP96,WyssGR01,HuhtalaKPT99}.  There are two important differences between these approaches and our work: 1) instead of discovering the FDs from scratch, we start with a set of provided FDs which have a certain level of trust, and we aim for a minimal modification of the provided FDs that yields at most $\tau$ violations; 2) in previous work, there are only ``local'' guarantees on the goodness (i.e., the number of violating tuples) of each FD, whereas in this paper we must make ``global'' guarantees that the whole set of FDs cannot be violated by more than $\tau$ tuples.  Thus, existing techniques for FD discovery are not applicable to our problem.

\section{Conclusions}
\label{sec:conclusion}

In this paper, we studied a data quality problem in which we are given a data set that does not satisfy the specified integrity constraints (namely, Functional Dependencies), and we are uncertain whether the data or the FDs are incorrect.  We proposed a solution that computes various suggestions for how to modify the data and/or the FDs (in a nearly-minimal way) in order to achieve consistency.  These suggestions cover various points on the spectrum of relative trust between the data and the FDs, and can help users determine the best way to resolve inconsistencies in their data.
We believe that our relative trust framework is relevant and applicable to many other types of constraints, such as conditional FDs (CFDs), inclusion dependencies and logical predicates.  In future work, we plan to develop cleaning algorithms within our framework for these constraints.


\bibliographystyle{abbrv}

\appendix

\section{Proofs}
\label{sec:proofs}

\subsection{Proof of Theorem~\ref{thm:mapping}}
\label{sec:proof:mapping}

In the following, we prove the first part of the theorem. Let $(\Sigma',I')$ be a $\tau$-constrained repair. It follows that no repair $(\Sigma'',I'')$ has $(dist_c(\Sigma,\Sigma''),dist_d(I,I'')) \prec (dist_c(\Sigma,\Sigma'),\tau)$. Because $dist_d(I,I') \leq \tau$, there is no repair $(\Sigma'',I'')$ satisfies $(dist_c(\Sigma,\Sigma''),dist_d(I,I'')) \prec (dist_c(\Sigma,\Sigma'),dist_d(I,I'))$. In other words, $(\Sigma',I')$ is a minimal repair.

We prove the second part by contradiction. Assume that $(\Sigma',I')$ is a minimal repair but it is not a $\tau$-constrained repair for the values of $\tau$ described in Equation~\ref{eq:range}. Because $\tau \geq dist_d(I,I')$, and based on Definition~\ref{def:const_min}, there must exist a repair $(\Sigma_x,I_x)$ such that $(dist_c(\Sigma,\Sigma_x),dist_d(I,I_x)) \prec (dist_c(\Sigma,\Sigma'),\tau)$ (if multiple repairs satisfy this criteria, we select the repair with the minimum distance to $I$, and we break ties using the smaller distance to $\Sigma$). Repair $(\Sigma_x,I_x)$ is a minimal repair because no other repair can dominate $(\Sigma_x,I_x)$ with respect to distances to $I$ and $\Sigma$. Because $(\Sigma',I')$ is a minimal repair, then $dist_d(I,I_x) \geq dist_d(I,I')$ (otherwise, $(\Sigma',I')$ would be dominated by $(\Sigma_x,I_x)$). However, existence of $(\Sigma_x,I_x)$ contradicts the fact that no minimal repair exists with distance to $I$ in the range $(dist_d(I,I'),\tau)$ (based on the value of $\tau$ obtained by Equation~\ref{eq:range}).

\subsection{Proof of Theorem~\ref{thm:tau_repairs}}
\label{sec:proof:tau_repairs}

For a generated repair $(\Sigma',I')$, the condition $dist_d(I,I') \leq \tau$ holds due to the constraint $\delta_{opt}(\Sigma',I) \leq \tau$ in line 1. For any $\Sigma'' \in \mathcal{S}(\Sigma)$, $\delta_{opt}(\Sigma'',I) \leq dist_d(I,I'')$ for all $I'' \models \Sigma''$, and thus the condition $dist_d(I,I'') \leq \tau$ implies that $\delta_{opt}(\Sigma'',I) \leq \tau$. Therefore the condition $\nexists \Sigma'' \in \mathcal{S}(\Sigma) (\delta_{opt}(\Sigma'',I) \leq \tau \wedge dist_c(\Sigma,\Sigma'') < dist_c(\Sigma,\Sigma'))$ in line 1, along with the tie breaking mechanism, imply that $\nexists  (\Sigma'',I'')\in {\bf U} \ ( dist_c(\Sigma,\Sigma''), dist_d(I,I'')) \prec (dist_c(\Sigma,\Sigma'),\tau)$. Thus, $(\Sigma',I')$ is a $\tau$-constrained repair.

\subsection{Proof of Lemma~\ref{lem:lower_bound}}
\label{sec:proof:lower_bound}

Let $\Sigma$ be the set of FDs corresponding to $S$. Assume that we are using the entire set of difference sets, denoted $\mathcal{D}_{all}$, that violate $\Sigma$ rather than using a subset of difference sets (line 13 in Algorithm~\ref{alg:astar}).

The cheapest goal state $S_g$ that are a descendent of $S$ will be among the states returned by the procedure \texttt{getDescGoalStates} because the procedure \texttt{getDescGoalStates} returns all minimal goal states (if any), and $S_g$ is minimal (i.e., there exist no other state $S'$ such that $S_g$ extends $S'$ and $S'$ is a goal state). 

Because we are using a subset of all difference sets $\mathcal{D}_{all}$, the cost of the reported cheapest goal state is less than or equal to the actual cost of the cheapest goal state. 

\subsection{Proof of Lemma~\ref{lem:find_assignment}}
\label{sec:proof:find_assignment}

We first prove the soundness of the algorithm. That is, we need to prove that if a tuple $t_c$ is returned, $t_c[A]=t[A]$, for $A\in Fixed\_Attrs$, and for all $t' \in I' \setminus C_{2opt}$, $(t_c,t')$ do not violate $\Sigma$.

From the algorithm description, it is clear that the condition $t_c[A]=t[A]$ holds, for $A\in Fixed\_Attrs$. Also, the condition in line 2 ensures that whenever a tuple $t_c$ is returned, there does not exist $t' \in I' \setminus C_{2opt}$ such that $(t_c,t')$ violate any FD in $\Sigma$.

We prove the completeness of the algorithm by contradiction. Assume that Algorithm~\ref{alg:find_assignment} returns $\phi$, while there exist a tuple $t_g$ that satisfies the conditions  $t_g[A]=t[A]$, for $A\in Fixed\_Attrs$, and $(t_g,t')$ satisfy $\Sigma$, for all $t' \in I' \setminus C_{2opt}$.

We first show that, just before returning $\phi$ at line 4, all attributes in $Fixed\_Attrs$ have equal values in tuples $t_c$ and $t_g$. This is clearly true for the initial value of $Fixed\_Attrs$. Let $A$ be the attribute that is first inserted in $Fixed\_Attrs$ in line 7. Before setting $t_c[A]$ to $t'[A]$ in line 6, there exist a tuple $t'\in I'\setminus C_{2opt}$ such that $(t',t_c)$ violate an FD $X \rightarrow A\in \Sigma$. Attributes in $X$ belong to $Fixed\_Attrs$ because attributes outside $Fixed\_Attrs$ are assigned to new variables (line 1) and cannot be equal to attributes of any other tuples. It follows that attribute $A$ in any valid solution must be equal to $t'[A]$ in order to satisfy $\Sigma$. Thus, $t_c[A]=t_g[A] =t'[A]$. The same argument is valid for attributes that are successively inserted into $Fixed\_Attrs$ before returning $\phi$. When the algorithm returns $\phi$ (line 4), there exists a tuple $t'\in I'\setminus C_{2opt}$ such that $(t',t_c)$ violate an FD $X \rightarrow A\in \Sigma$ and attribute $A$ belongs to $Fixed\_Attrs$. Because $AX \subset Fixed\_Attrs$, and attributes in $Fixed\_Attrs$ have equal values in $t_c$ and $t_g$, it follows that $(t',t_g)$ violate $X \rightarrow A$ as well (i.e., $t_g$ is not a valid answer), which contradicts our initial assumption.

\subsection{Proof of Theorem~\ref{thm:P}}
\label{sec:proof:P}

We first prove that the returned $I'$ satisfies $\Sigma'$. Let $G$ be the conflict graph of $I$ with respect to $\Sigma'$ and let $C_{2opt}$ be a 2-approximate minimum vertex cover of $G$ that is obtained at line 2 in Algorithm~\ref{alg:get_instance_repair}. The tuple set $I \setminus C_{2opt}$ satisfies $\Sigma'$, and thus the corresponding tuples in $I'$ satisfy $\Sigma'$ as well. For each tuple $t$ that is randomly picked from $C_{2opt}$ in line 5 in Algorithm~\ref{alg:get_instance_repair}, modifying $t$ as described in lines 6-15 makes the set $I' \setminus C_{2opt} \cup \{t\}$ satisfies $\Sigma'$, as we show in the following. We observe that for a $Fixed\_Attrs$ containing a single attribute $A$, there exists an assignment to the attributes $R \setminus \{A\}$ in $t$ such that $I' \setminus C_{2opt} \cup \{t\}$ satisfies $\Sigma'$ (i.e., $t_c$ cannot be $\phi$ at line 7). We describe one possible assignment as follows. If the value of $t[A]$ does not appear in attribute $A$ of any tuple in $I' \setminus C_{2opt}$, then setting attributes $R \setminus \{A\}$ to new variables is a valid assignment. Otherwise, let $t_r$ be a tuple in $I' \setminus C_{2opt}$ such that $t[A] =t_r[A]$. Setting attributes $R\setminus \{A\}$ in $t$ to the values of corresponding attributes in $t_r$ is a valid assignment. Thus, $t_c$ cannot be $\phi$ in line 7 due to completeness of Algorithm~\ref{alg:find_assignment}, which is proved in Lemma~\ref{lem:find_assignment}.

After each iteration of the while loop in line 8, Algorithm~\ref{alg:get_instance_repair} maintains a tuple $t_c$ such that current attributes in $Fixed\_Attrs$ have equal values in $t_c$ and the current version of $t$, and other attributes outside $Fixed\_Attrs$ in $t_c$ are assigned to values that make $I' \setminus C_{2opt} \cup \{t_c\}$ satisfies $\Sigma'$ (due to soundness of Algorithm~\ref{alg:find_assignment} as proved in Lemma~\ref{lem:find_assignment}). After inserting all attributes in $Fixed\_Attrs$, $t$ is equal to $t_c$ and thus $I' \setminus C_{2opt} \cup \{t\}$ satisfies $\Sigma'$. After processing, and removing, all tuples from $C_{2opt}$, the resulting instance $I'$ satisfies $\Sigma'$.

We prove the approximate optimality of the algorithm as follows. Let $C_{opt}$ be a minimum vertex cover of $G$. The minimum number of cell changes $\delta_{opt}(\Sigma',I)$ must be greater than or equal to $|C_{opt}|$. This can be proved by contradiction as follows. Assume that there exists an instance $I' \models \Sigma'$ such that the number of changed cells in $I'$ is less than $|C_{opt}|$. Let $T$ be the set of changed tuples in $I'$. $T$ represents a vertex cover of $G$ and $|T| < |C_{opt}|$, which contradicts minimality of $C_{opt}$. 

In the following, we prove that the number of changed cells is $|C_{2opt}| \cdot \min\{|R|-1,|\Sigma|\}$, which is $2 \cdot \min\{|R|-1,|\Sigma|\}$-approximate minimum, based on the fact that $\delta_{opt}(\Sigma',I) \geq |C_{opt}|$. The algorithm changes only attributes of tuples in $C_{2opt}$. Furthermore, we prove that the number of changed cells in each tuple in $C_{2opt}$ is at most $\min\{|R|-1,|\Sigma|\}$. It is clear that the maximum number of changed cells in each tuple is $|R|-1$ because the first attribute inserted into $Fixed\_Attrs$ cannot be changed (line 6 in Algorithm~\ref{alg:get_instance_repair}).

We show that after changing $|\Sigma'|$ attributes in $t$, the set $I' \setminus C_{2opt} \cup \{t\}$ satisfies $\Sigma'$ and thus no other attributes in $t$ need to be changed. In general, we prove that after the $k$-th change to $t$, $I' \setminus C_{2opt} \cup \{t\}$ can violate at most $|\Sigma'| - k$ FDs in $\Sigma'$. Let $B$ be a changed attribute in $t$. If $B$ was changed to a variable, there must exist an FD $X\rightarrow A \in \Sigma'$ such that $B \in X$. The reason is that if $B$ does not appear in any FD, it cannot be changed by Algorithm~\ref{alg:get_instance_repair}, and if $B$ only appears as a right-hand-side attribute in FDs in $\Sigma'$, it can only remain unchanged or be changed to a constant. It follows that $(t,t')$ cannot violate $X \rightarrow A$, for all $t' \in I'\setminus C_{2opt}$ after changing $t[B]$ to a variable and adding $B$ to $Fixed\_Attrs$. If $B$ was changed to a constant, there must exist an FD $X \rightarrow B \in \Sigma'$ and another FD $Y \rightarrow X$ implied by $\Sigma'$ such that $Y \subset Fixed\_Attrs$ and values of attributes in $Y$ are constants (refer to lines 2-9 in Algorithm~\ref{alg:find_assignment}). In successive iterations, attributes in $X$ cannot be assigned to values other than the current constants in $t_c$, otherwise $Y \rightarrow X$ would be violated. It follows that $(t,t')$ cannot violate $X \rightarrow B$, for $t' \in I'\setminus C_{2opt}$ in successive iterations. After changing $|\Sigma'|$ attributes in $t$, we do not need to perform further changes. Because $|\Sigma'| \leq |\Sigma|$, it follows that the maximum number of attributes changed for each tuple in $C_{2opt}$ is $|\Sigma|$, which completes the proof.

\end{document}